\documentclass[pdflatex,sn-mathphys-num]{sn-jnl}

\usepackage{caption}%
\usepackage{subcaption}%
\usepackage{multicol}%

\usepackage{geometry}
\usepackage{xcolor}
\usepackage{enumitem}
\usepackage{graphicx}
\usepackage{mathtext}
\usepackage{mathrsfs}
\usepackage{amsmath} 
\usepackage{hyperref}
\usepackage{booktabs} 
\usepackage{multirow} 
\usepackage{makecell}

\usepackage{setspace}
\begin{document}


\title{CKM determination from $W$ decays with jet flavor tagging at CEPC\protect\\
\vspace{1em}\normalfont\normalsize\color{black!70} Date: \today}

\author[1]{\fnm{Zhao-Ling} \sur{Zhang}}
\author[2]{\fnm{Li-Bo} \sur{Liao}}
\author[3]{\fnm{Jia-Rong} \sur{Li}}
\author[4,5]{\fnm{Jia-Bao} \sur{Gong}}
\author[1]{\fnm{Wei-Min} \sur{Song}}
\author[4,5]{\fnm{Gang} \sur{Li}}

\affil[1]{\orgdiv{College of Physics}, \orgname{Jilin University}, \orgaddress{\city{Changchun}, \postcode{130012}, \country{China}}}
\affil[2]{\orgdiv{School of Science}, \orgname{Sun Yat-sen University}, \orgaddress{\city{Shenzhen}, \postcode{518107}, \country{China}}} 
\affil[3]{\orgname{Tsinghua University}, \orgaddress{\city{Beijing}, \postcode{100084}, \country{China}}}
\affil[4]{\orgdiv{Institute of High Energy Physics}, \orgname{Chinese Academy of Sciences}, \orgaddress{\city{Beijing}, \postcode{100049}, \country{China}}}
\affil[5]{\orgname{University of Chinese Academy of Sciences}, \orgaddress{\city{Beijing}, \postcode{100049}, \country{China}}}

\abstract{This study investigates the sensitivity of the CEPC to CKM matrix elements $|V_{ij}|$ in semileptonic $W$-pair decays using simulated $e^+e^- \to W^+W^- \to \mu\nu \bar{q}q^\prime$ samples corresponding to $21.6\,\mathrm{ab}^{-1}$ at $\sqrt{s}=240\,\mathrm{GeV}$, with $|V_{ub}|$ not extracted because of the negligible expected $\bar{u}b$ yield.
The projected statistical precisions reach $0.59\,\%$ for $|V_{cb}|$ and $0.01\,\%$ for $|V_{cs}|$, indicating that the CEPC can provide direct, high-precision, and largely model-independent determinations of CKM matrix elements from hadronic $W$ decays.
The main sources of systematic uncertainty are discussed, and further improvements are expected from both experimental and theoretical developments, including dedicated data-driven calibrations of detector response and jet flavor tagging performance, as well as more precise higher-order calculations of the relevant electroweak and QCD corrections. Such measurements would provide stringent consistency tests of the Standard Model charged-current flavor structure and offer sensitivity to possible new physics effects through deviations from the Standard Model expectations.}
\keywords{CKM matrix, CEPC, $W$ boson decays, ParticleNet}

\maketitle

\section{Introduction}

The Cabibbo-Kobayashi-Maskawa (CKM) matrix is a central ingredient of the Standard Model (SM), describing quark flavor mixing in charged-current weak interactions. Precise determinations of its elements are essential for testing CKM unitarity and probing physics beyond the SM. According to the latest \textit{Review of Particle Physics} by the Particle Data Group (PDG)~\cite{pdg2024}, $|V_{ud}|$ and $|V_{us}|$ are already known with very high precision from nuclear beta decays~\cite{Hardy:2020qwl} and kaon decays~\cite{Antonelli:2010yf}. By contrast, the determination of $|V_{cd}|$, $|V_{cs}|$, and especially $|V_{cb}|$ still relies mainly on semileptonic decays of $D$ and $B$ mesons, and therefore on theoretical input from lattice QCD for the relevant hadronic form factors~\cite{Aoki:2021kgd}. The long-standing tension between inclusive and exclusive determinations of $|V_{cb}|$ further illustrates the difficulty of controlling these non-perturbative effects~\cite{pdg2024}.

Hadronic decays of the $W$ boson provide a theoretically clean alternative for extracting CKM matrix elements.
In the SM, the partial width for $W^- \to \bar{q}_i q^\prime_j$ can be written as~\cite{CMS_W_branching_2022}
\begin{equation}
\Gamma(W^- \to \bar{q}_i q^\prime_j) = \frac{\sqrt{2} G_F N_c}{12\pi}m_W^3 |V_{ij}|^2 \left[ 1 + \sum_{k=1}^4 c^{(k)}_{\rm QCD} \left(\frac{\alpha_S}{\pi}\right)^k + \delta_{\rm EW}(\alpha) + \delta_{\rm mix}(\alpha \alpha_S) \right],
\label{eq:partial_width}
\end{equation}
where the prefactor is the Born-level width and the terms in brackets denote the higher-order QCD, electroweak, and mixed corrections.
Unless otherwise stated, charge-conjugate processes are implied throughout this paper.
The QCD coefficients $c_{\rm QCD}^{(k)}$ are known up to $k=4$~\cite{Baikov:2008jh}, the electroweak correction $\delta_{\rm EW}$ is known to ${\cal O}(\alpha)$~\cite{Denner:1992vx}, and the mixed correction $\delta_{\rm mix}$ is known to ${\cal O}(\alpha\alpha_S)$~\cite{Kara:2013dua}. Since these corrections are flavor independent, they cancel to a very good approximation in ratios of partial widths. The exclusive branching fraction can therefore be expressed as
\begin{equation}
{\rm BR}(W^- \to \bar{q}_i q^\prime_j) = \frac{|V_{ij}|^2}{\sum_{i' \in \{u,c\}} \sum_{j' \in \{d,s,b\}} |V_{i'j'}|^2}.
\label{eq:br}
\end{equation}
Assuming CKM unitarity, so that $\sum_{i' \in \{u,c\}} \sum_{j' \in \{d,s,b\}} |V_{i'j'}|^2 = 2$, one obtains ${\rm BR}(W^- \to \bar{q}_i q^\prime_j) = \frac{1}{2}|V_{ij}|^2$. This relation enables a direct extraction of $|V_{ij}|$ from hadronic $W$ branching fractions without hadronic form factors.

Experimentally, however, this program remains challenging because it requires the identification of specific quark flavors in hadronic final states, including light-flavor separation. At hadron colliders, this is strongly limited by large combinatorial backgrounds and the complexity of the experimental environment. LEP provided a much cleaner $e^+e^-$ collision environment with well-defined initial-state kinematics, making direct studies of hadronic $W$ decays possible. However, the precision of such measurements was restricted by the available $W^+W^-$ statistics and by the jet flavor tagging techniques available at the time~\cite{LEP2_Vcs}. Future $e^+e^-$ colliders, including CEPC~\cite{cepc_cdr}, FCC-ee~\cite{fcc_ee_cdr}, ILC~\cite{ilc_tdr}, and CLIC~\cite{clic_summary}, provide a particularly favorable environment in which to revisit this problem. Direct determinations of $|V_{cs}|$ and $|V_{cb}|$ from $W$ decays using jet flavor tagging have been studied for FCC-ee~\cite{Marzocca_CKM_Wdecays}, and measurements of $|V_{cb}|$ in semileptonic $W^+W^-$ events have been investigated at future Higgs factories~\cite{Liang_Vcb_HiggsFactory}. In particular, operating at $\sqrt{s}=240$\,GeV with an integrated luminosity of 21.6\,ab$^{-1}$, the CEPC is expected to deliver a very large $W^+W^-$ sample in a clean experimental environment. At the same time, recent progress in machine-learning-based jet flavor tagging, notably with ParticleNet~\cite{Qu:2019gqs} and ParticleTransformer (ParT)~\cite{qu2022particle}, directly improves the separation of $b$, $c$, $s$~\cite{Kats_strange_tagging}, and light-flavor jets. Jet-origin identification has also demonstrated the ability to distinguish different quark flavors and to separate quark from antiquark jets at an $e^+e^-$ Higgs factory~\cite{Liang_JetOrigin}. This improved multi-class flavor discrimination is crucial for extracting exclusive hadronic $W$ branching fractions and therefore substantially enhances the sensitivity to CKM matrix elements from hadronic $W$ decays.

This study investigates the sensitivity of the CEPC to five CKM matrix elements, $|V_{ud}|$, $|V_{us}|$, $|V_{cd}|$, $|V_{cs}|$, and $|V_{cb}|$, using the semileptonic channel $e^+e^- \to W^+W^- \to \mu\nu \bar{q}q^\prime$. For brevity, $\ell\nu$ denotes the corresponding charged-lepton--neutrino final state in the explicit process labels below; for example, $\mu\nu \equiv \mu^- \bar{\nu}_\mu$, $e\nu \equiv e^- \bar{\nu}_e$, and $\tau\nu \equiv \tau^- \bar{\nu}_\tau$. The small $\bar{u}b$ component is retained in the template fit with fixed normalization, while $|V_{ub}|$ is not extracted because its expected yield is negligible. The $\mu\nu \bar{q}q^\prime$ final state provides a clean topology with an isolated muon and two jets, enabling the hadronic $W$ decay to be reconstructed with high purity. 
Based on realistic detector simulation and ParticleNet-based jet flavor tagging, a simultaneous template fit to the event-level flavor discriminants, referred to below as the template fit, is performed to extract the corresponding branching fractions and CKM matrix elements.

This paper is organized as follows. Section~2 describes the CEPC detector simulation and the signal and background Monte Carlo samples. Section~3 presents the event selection and background study. Section~4 introduces the jet flavor tagging framework based on ParticleNet. Section~5 describes the template fit and the projected precision on the CKM matrix elements. Section~6 summarizes the main results.

 
\section{CEPC Detector and Simulation Samples}

The CEPC is a circular $e^+e^-$ collider with a circumference of 100\,km, designed to operate at several center-of-mass energies, including the $Z$ pole (91\,GeV), the $WW$ threshold (161\,GeV), and the Higgs-factory energy (240\,GeV)~\cite{cepc_tdr_acc}. The CEPC reference detector, AGORA~\cite{cepc_tdr_dec}, is designed for precision measurements and searches for new physics. Its tracking system consists of a six-layer silicon pixel vertex detector, an inner silicon tracker with three barrel layers and endcap disks, a time projection chamber, and an outer silicon tracker with timing capability. The calorimeter system is optimized for particle-flow reconstruction and comprises a crystal electromagnetic calorimeter and a steel--scintillating-glass sampling hadronic calorimeter. Both calorimeters are located inside the superconducting solenoid, with a muon detection system embedded in the surrounding iron return yoke. These systems provide precise track and vertex reconstruction and support jet reconstruction and flavor tagging.

Monte Carlo samples for both signal and background processes are generated at $\sqrt{s}=240$\,GeV using \textsc{Whizard}~1.9.5~\cite{whizard,whizard_arxiv}, with initial-state radiation (ISR) and bremsstrahlung effects included. Parton showering and hadronization are simulated with \textsc{Pythia}~6.4~\cite{pythia6}. Detector response is modeled using the parameterized \textsc{Delphes} fast-simulation framework~\cite{delphes1,delphes2}, configured according to the CEPC reference detector design~\cite{cepc_tdr_dec}. The tracking response is parameterized using the \texttt{TrackCovariance} module~\cite{delphes_trackcovariance}, which evaluates the track-parameter covariance matrix for each track. All simulated samples are normalized to an integrated luminosity of 21.6\,ab$^{-1}$ at $\sqrt{s}=240$\,GeV~\cite{cepc_tdr_acc}. The cross sections and expected event yields are summarized in Table~\ref{table1}. Backgrounds from two-photon processes are neglected in this analysis. Although their production cross sections are large, they typically have very low visible energy and are efficiently removed by the basic kinematic selections.

This study focuses on the $W^+W^- \to \mu\nu \bar{q}q^\prime$ final state, which provides a clean experimental signature and a straightforward event-selection strategy. Compared with the $W^+W^- \to e\nu \bar{q}q^\prime$ channel, the muon channel is much less affected by single-$W$ and single-$Z$ backgrounds. In the electron channel, these backgrounds are strongly enhanced by $t$-channel contributions, making them topologically irreducible and requiring a more involved treatment of interference effects with the signal. The signal considered here consists of six exclusive $e^+e^- \to W^+W^- \to \mu\nu \bar{q}_i q^\prime_j$ channels, in which one $W$ boson decays leptonically and the other decays hadronically into a specific quark pair. The six flavor combinations are $\bar{u}d$, $\bar{u}s$, $\bar{u}b$, $\bar{c}d$, $\bar{c}s$, and $\bar{c}b$, each probing the corresponding CKM factor $|V_{ij}|^2$.
The expected event yields are derived using ${\rm BR}(W^- \to \mu^- \bar{\nu}_{\mu})=10.63\,\%$ from the PDG~\cite{pdg2024} and the six exclusive hadronic branching fractions from Ref.~\cite{arxiv_W_branching_fractions}, which are listed in Table~\ref{tab:w_hadronic_br}.

\begin{table}[!htbp]
\centering
\caption{Monte Carlo samples for the signal and background processes at $\sqrt{s}=240$\,GeV and an integrated luminosity of 21.6\,ab$^{-1}$, together with the corresponding cross sections and expected event yields. The notation $up$ denotes $u$, $\bar{u}$, $c$, and $\bar{c}$ quarks, while $down$ denotes $d$, $\bar{d}$, $s$, $\bar{s}$, $b$, and $\bar{b}$ quarks.}
\label{table1}
\begin{tabular}{lrr}
\toprule
\textbf{Processes} & \textbf{Cross section (fb)} & \textbf{Expected events} \\
\midrule
\multicolumn{3}{l}{Signal: \text{$e^+e^-\to W^+W^-\to \mu\nu \bar{q}q^\prime$}} \\

$W^+W^-\to \mu\nu \bar{u}d$ &   1173.02   & 25,337,232 \\
$W^+W^-\to \mu\nu \bar{u}s$ &   62.71     & 1,354,536  \\
$W^+W^-\to \mu\nu \bar{u}b$ &   0.02      & 432        \\
$W^+W^-\to \mu\nu \bar{c}d$ &   62.71     & 1,354,536  \\
$W^+W^-\to \mu\nu \bar{c}s$ &   1169.34   & 25,257,744 \\
$W^+W^-\to \mu\nu \bar{c}b$ &   2.18      & 47,088     \\
\midrule
\multicolumn{3}{l}{\text{Background: 2-fermion}} \\

$e^+e^-\to q\bar{q}$ & 54,106.86 & 1,168,708,160 \\
$e^+e^-\to \ell^+\ell^-$ & 34,856.50 & 752,900,400 \\
$e^+e^-\to \nu\bar{\nu}$ & 54,099.51 & 1,168,549,360 \\
\midrule
\multicolumn{3}{l}{\text{Background: 4-fermion ($ZZ$)}} \\

$ZZ\to \mu^+\mu^-\,\mathrm{up}\,\mathrm{up}$ & 87.39 & 1,887,624 \\
$ZZ\to \mu^+\mu^-\,\mathrm{down}\,\mathrm{down}$ & 136.14 & 2,940,624 \\
$ZZ\to \nu\bar{\nu}\,\mathrm{up}\,\mathrm{up}$ & 84.38 & 1,822,608 \\
$ZZ\to \nu\bar{\nu}\,\mathrm{down}\,\mathrm{down}$ & 139.71 & 3,017,736 \\
$ZZ\to \tau^+\tau^-\,\mathrm{up}\,\mathrm{up}$ & 41.56 & 897,696 \\
$ZZ\to \tau^+\tau^-\,\mathrm{down}\,\mathrm{down}$ & 67.31 & 1,453,896 \\
\midrule
\multicolumn{3}{l}{\text{Background: 4-fermion ($W^+W^-$)}} \\

$W^+W^-\to \tau\nu\,\mathrm{up}\,\mathrm{down}$ & 2662.03 & 57,499,848 \\
$W^+W^-\to e\nu\,\mathrm{up}\,\mathrm{down}$    & 2505.30 & 54,114,480 \\
\midrule
\multicolumn{3}{l}{\text{Background: 4-fermion (single $Z$)}} \\
$Z\to e^+e^-\,\mathrm{up}\,\mathrm{up}$ & 190.21 & 4,108,536 \\
$Z\to e^+e^-\,\mathrm{down}\,\mathrm{down}$ & 125.83 & 2,717,928 \\
$Z\to \nu\bar{\nu}\,\mathrm{up}\,\mathrm{up}$ & 55.59 & 1,200,744 \\
$Z\to \nu\bar{\nu}\,\mathrm{down}\,\mathrm{down}$ & 90.03 & 1,944,648 \\
\midrule
\multicolumn{3}{l}{\text{Background: 4-fermion (single $W$)}} \\

$e^+e^-\to e\nu\mu\nu$ & 436.70 & 9,432,720 \\
$e^+e^-\to e\nu\tau\nu$ & 435.93 & 9,416,088 \\
$e^+e^-\to e\nu\,\mathrm{up}\,\mathrm{down}$ & 2,612.62 & 56,432,592 \\
\midrule
\multicolumn{3}{l}{\text{Background: $ZH$}} \\

$ZH\to \ell^+\ell^- H$ & 20.56 & 444,096 \\
$ZH\to \nu\bar{\nu} H$ & 46.29 & 1,000,064 \\
$ZH\to q\bar{q} H$ & 136.81 & 2,955,096 \\
\bottomrule
\end{tabular}
 \end{table}

The SM backgrounds are grouped according to their production mechanism and event topology, following Ref.~\cite{moxin_w_physics}. The two-fermion ($2f$) processes, including $e^+e^- \to q\bar{q}$, $e^+e^- \to \ell^+\ell^-$, and $e^+e^- \to \nu\bar{\nu}$, have the largest production cross sections. However, their relatively simple final-state topologies differ substantially from that of the signal, so their contributions become negligible after event selection. The dominant surviving backgrounds arise from four-fermion ($4f$) processes, which more easily mimic the signal topology. This category includes $ZZ$ production with semi-visible final states, as well as single-$W$ and single-$Z$ production. In addition, $W^+W^-$ events with $W^- \to e^- \bar{\nu}_e$ or $W^- \to \tau^- \bar{\nu}_{\tau}$ contribute when electrons are misidentified or when $\tau$ decays produce muons. Higgs-associated production ($ZH$) is also included, since specific decay modes can lead to similar final states.

 \begin{table}[htbp]
\centering
\caption{SM predictions for the exclusive hadronic branching fractions of the $W$ boson~\cite{arxiv_W_branching_fractions}.}
\label{tab:w_hadronic_br}
\renewcommand{\arraystretch}{1.3}
\begin{tabular}{cc}
\toprule
Decay mode ($W^- \to \bar{q}_i q^\prime_j$) & Branching fraction \\
\midrule
$\bar{u}d$ & $31.8\,\%$ \\
$\bar{c}s$ & $31.7\,\%$ \\
$\bar{u}s$ & $1.7\,\%$ \\
$\bar{c}d$ & $1.7\,\%$ \\
$\bar{c}b$ & $5.9 \times 10^{-4}$ \\
$\bar{u}b$ & $4.5 \times 10^{-6}$ \\
\bottomrule
\end{tabular}
\end{table}

\section{Event Selection}

This section describes the event selection used to isolate the semileptonic $W^+W^- \to \mu\nu \bar{q}q^\prime$ signal from background processes at $\sqrt{s}=240$\,GeV. The signal topology is characterized by an energetic isolated muon, missing momentum carried by the neutrino, and two hadronic jets originating from the hadronic $W$ decay. A topology-driven cut-based selection is therefore adopted, in which backgrounds are suppressed sequentially according to their kinematic differences from the signal.

\begin{figure}[!htb]
    \centering
    \includegraphics[width=0.8\textwidth]{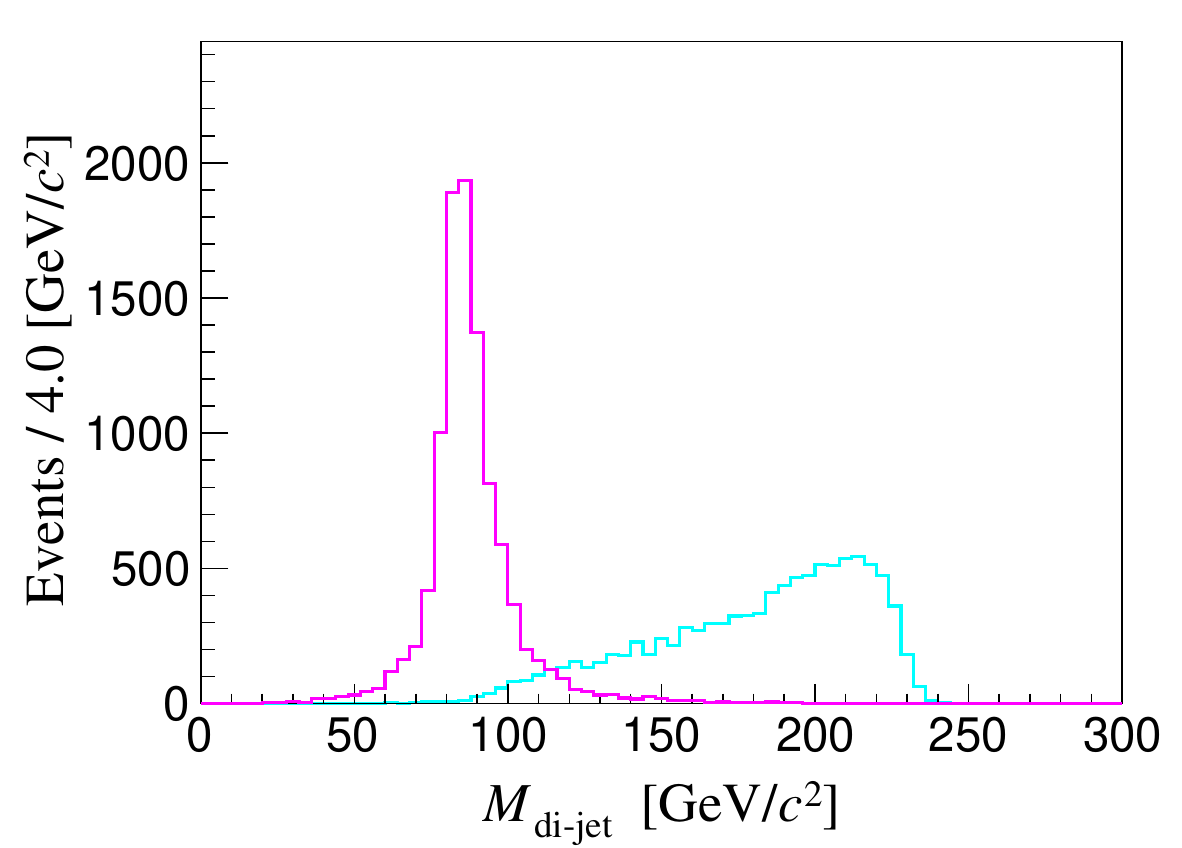}
    \caption{Reconstructed di-jet invariant-mass distributions, $M_{\text{di-jet}}$, for the signal process $W^+W^- \to \mu\nu \bar{c}s$. The blue and pink curves correspond to the distributions before and after removing the leading muon candidate prior to jet clustering, respectively.}
    \label{fig:m2jet_comparison}
\end{figure}

All events are reconstructed with the $e^{+}e^{-} k_t$ algorithm~\cite{eekt} implemented in \textsc{FastJet}~\cite{fastjet}, with the visible final state clustered into exactly two jets. To prevent the signal muon from being clustered into the hadronic system, the leading muon candidate is removed from the particle collection before jet clustering. This procedure significantly improves the di-jet mass resolution and, as shown in Fig.~\ref{fig:m2jet_comparison} for the $W^+W^- \to \mu\nu \bar{c}s$ channel, restores the hadronic $W$-boson mass peak.

\subsection{Preliminary Event Selection}

\begin{figure}[!htb]
    \centering
    \includegraphics[width=0.8\textwidth]{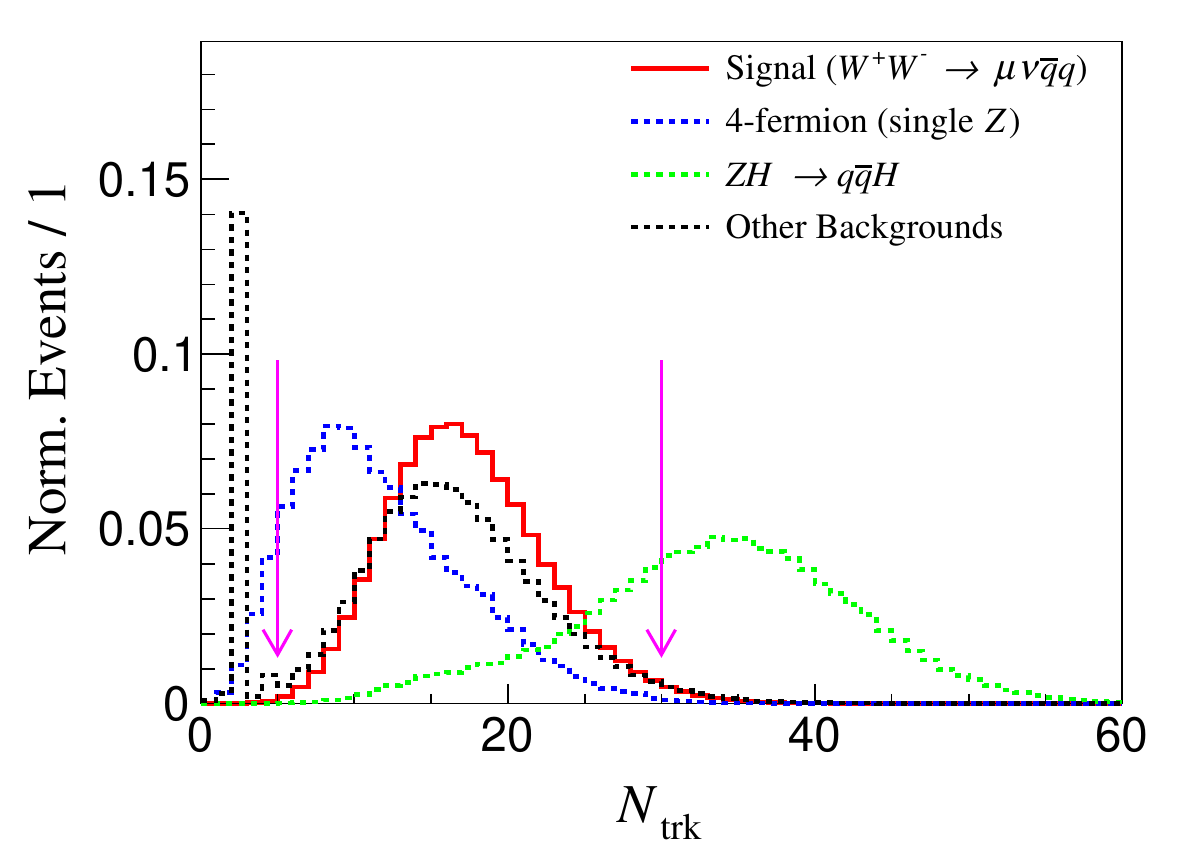}
    \caption{Normalized charged-track multiplicity distributions for the $W^+W^- \to \mu\nu \bar{q}q^\prime$ signal (red solid line), the single-$Z$ background (blue dashed line), the $ZH \to q\bar{q}H$ background (green dashed line), and the remaining combined backgrounds (black dashed line) at $\sqrt{s}=240$\,GeV.}
    \label{fig:Ncharge}
\end{figure}

Baseline event-level requirements are first imposed to suppress backgrounds with incompatible track multiplicities or global kinematic properties. Events are required to satisfy $5 \leq N_{\text{trk}} \leq 30$ and $50 \leq \sum E_{\text{trk}} \leq 200$\,GeV. The lower bounds reject low-multiplicity purely leptonic events such as $e^+e^- \to \ell^+\ell^-$, while the upper bounds reduce high-multiplicity fully hadronic backgrounds. To suppress ISR-dominated and collinear beam backgrounds, the leading-photon momentum is required to satisfy $P(\gamma_{\text{lead}}) < 60$\,GeV. In addition, at least one reconstructed lepton is required, $N_{\ell} \geq 1$, to suppress fully hadronic final states such as $W^+W^- \to \bar{q}q^\prime\bar{q}q^\prime$, $ZZ \to q\bar{q}q\bar{q}$, and $ZH \to q\bar{q}q\bar{q}$. The normalized charged-track multiplicity distributions that motivate the $N_{\text{trk}}$ requirement are shown in Fig.~\ref{fig:Ncharge}.

\subsection{Muon Identification and Isolation}

\begin{figure}[!htb]
    \centering
    \includegraphics[width=0.8\textwidth]{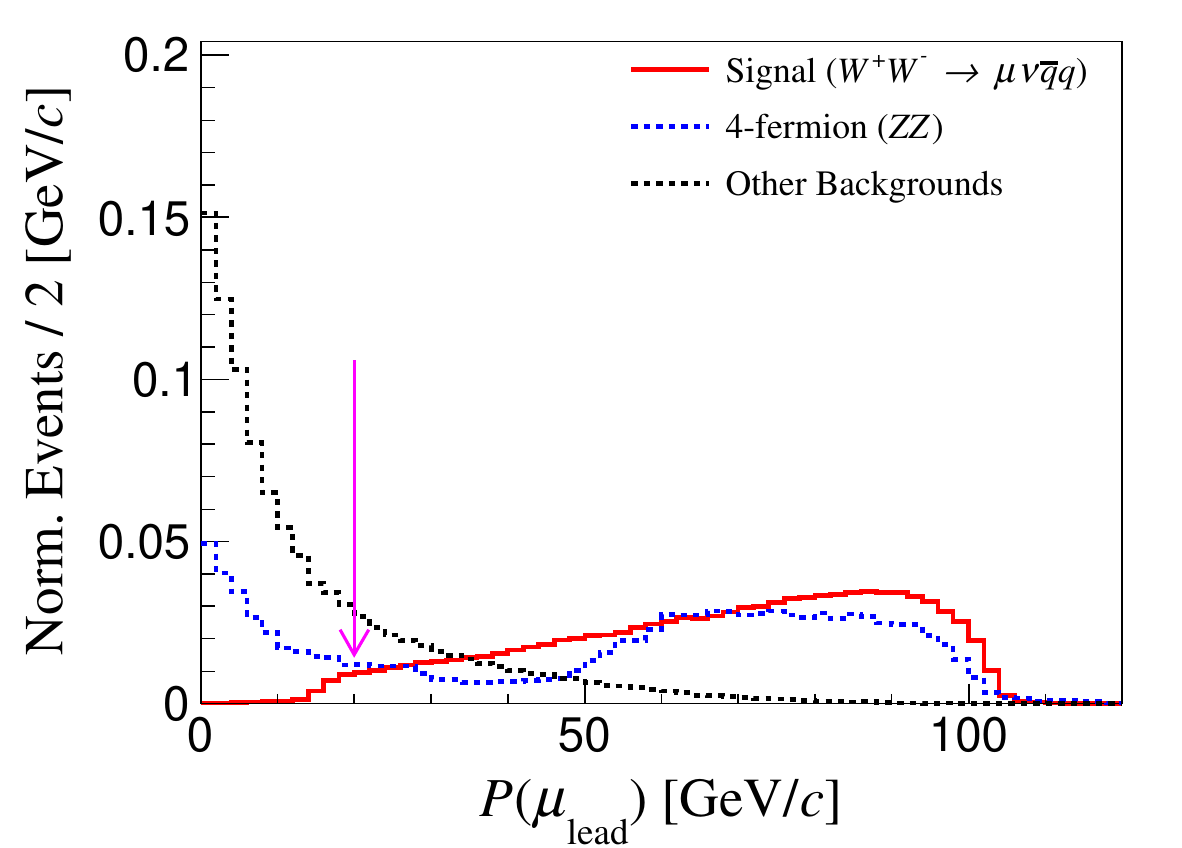}
    \caption{Normalized distributions of $P(\mu_{\text{lead}})$ after the preceding selection requirements. The signal is shown by the red solid line, the semileptonic $ZZ$ background by the blue dashed line, and the remaining combined backgrounds by the black dashed line. The pink arrow indicates the requirement $P(\mu_{\text{lead}}) \geq 20$\,GeV.}
    \label{fig:muon_vars}
\end{figure}

The defining signature of the signal is an energetic isolated muon originating from the leptonic decay of a $W$ boson. Events are therefore required to contain at least one reconstructed muon, $N_{\mu} \geq 1$. To ensure that this muon is prompt and to suppress soft muons from heavy-flavor hadron decays, the leading-muon momentum must satisfy $P(\mu_{\text{lead}}) \geq 20$\,GeV. Backgrounds containing multiple hard muons, such as $ZZ \to \mu^+\mu^- q\bar{q}$, are further suppressed by requiring the scalar sum of all muon momenta to satisfy $\sum P(\mu) \leq 100$\,GeV, consistent with the kinematics of a single leptonic $W$ decay at $\sqrt{s}=240$\,GeV. Since the signal muon recoils against the hadronic $W$ boson, a substantial angular separation from the di-jet system is additionally required, $80^\circ \leq \theta(\mu_{\text{lead}}, \text{di-jet}) \leq 170^\circ$. Finally, to reject residual QCD backgrounds with non-prompt muons embedded inside jets, a stringent isolation criterion is imposed: the energy ratio $E_{\text{ratio}}(\mu_{\text{lead}})$, defined as the fraction of visible energy within a $10^\circ$ cone around the leading muon that is carried by the muon itself, must satisfy $E_{\text{ratio}}(\mu_{\text{lead}}) \geq 0.9$. The discriminating power of the leading-muon momentum requirement is illustrated in Fig.~\ref{fig:muon_vars}.

\subsection{Hadronic System}

\begin{figure}[!htbp]
    \centering
    \includegraphics[width=0.8\textwidth]{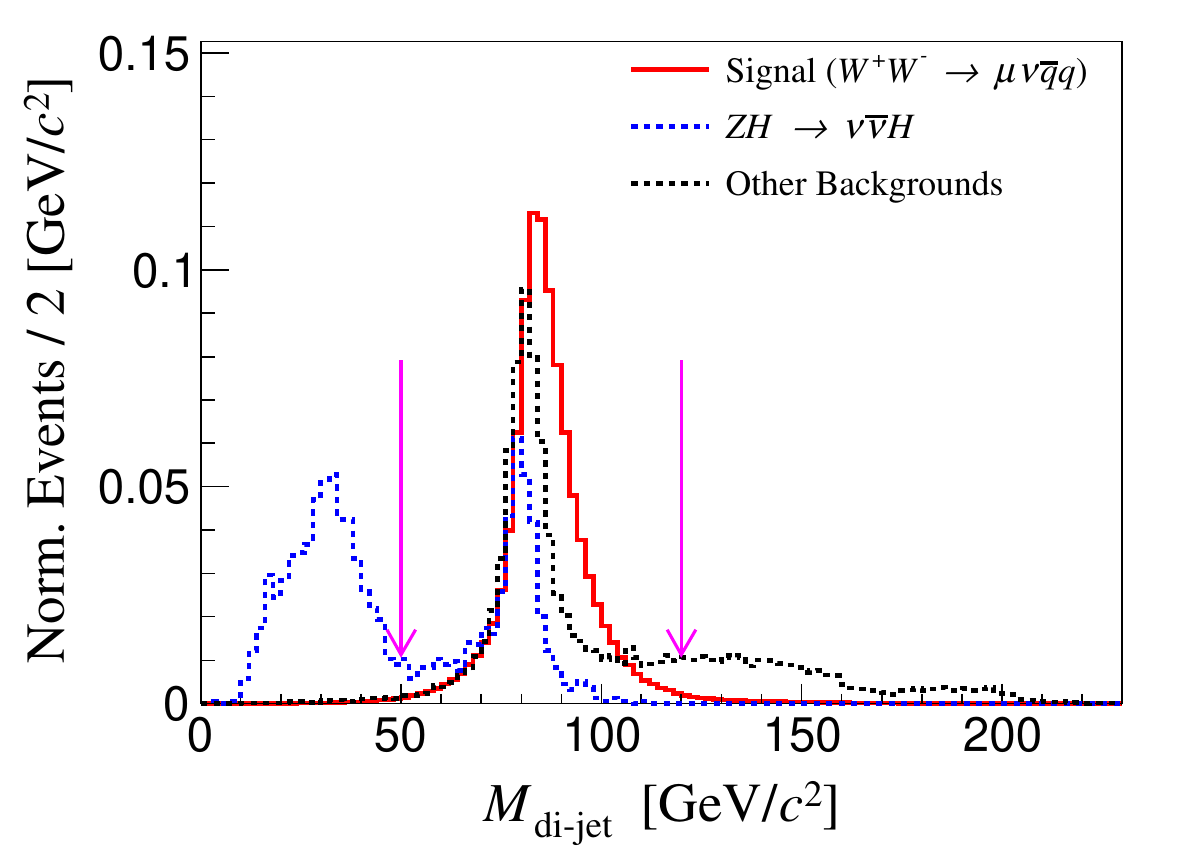}
    \caption{Normalized distributions of the di-jet invariant mass, $M_{\text{di-jet}}$, after the muon-identification and isolation requirements. The signal is shown by the red solid line, the $ZH \to \nu\bar{\nu}H$ background by the blue dashed line, and the remaining combined backgrounds by the black dashed line. The pink arrows indicate the requirement $50 \leq M_{\text{di-jet}} \leq 120$\,GeV.}
    \label{fig-va-M2jet}
\end{figure}

As described above, the hadronic system is reconstructed by clustering all visible particles into exactly two jets with the $e^{+}e^{-} k_t$ algorithm, explicitly excluding the leading muon from the input collection. The invariant mass of this di-jet system, $M_{\text{di-jet}}$, is the main observable used to identify the hadronically decaying $W$ boson. The requirement $50 \leq M_{\text{di-jet}} \leq 120$\,GeV is therefore applied, which efficiently retains the $W$ resonance while rejecting non-resonant backgrounds. The corresponding distributions are shown in Fig.~\ref{fig-va-M2jet}, where the signal is strongly concentrated around the nominal $W$-boson mass.

\begin{figure}[!htbp]
    \centering
    \includegraphics[width=0.8\textwidth]{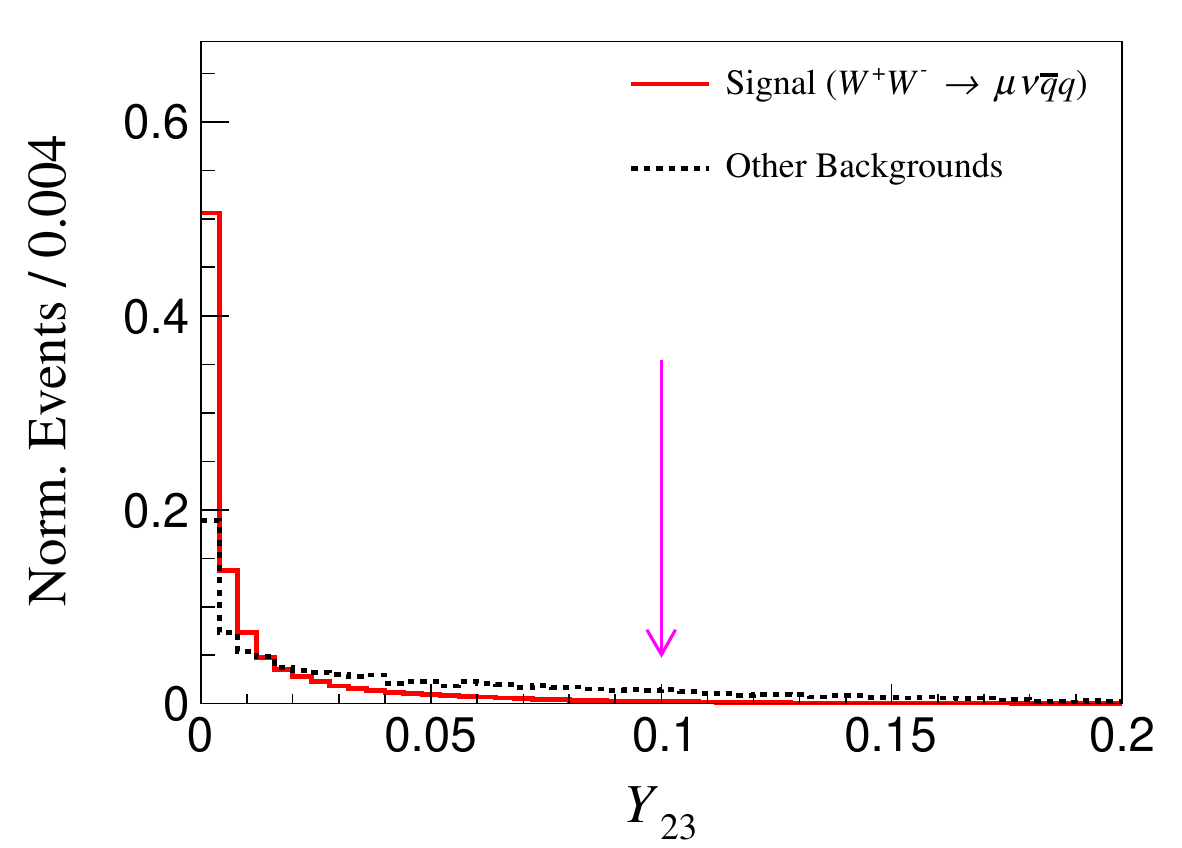}
    \caption{Normalized distributions of the jet-resolution parameter $Y_{23}$ after the $M_{\text{di-jet}}$ requirement. The signal is shown by the red solid line and the remaining combined backgrounds by the black dashed line. The pink arrow indicates the requirement $Y_{23} \leq 0.1$.}
    \label{fig:y23_var}
\end{figure}

Although the preceding requirements already remove most multi-jet backgrounds, a tail toward large $Y_{23}$ values remains, as shown in Fig.~\ref{fig:y23_var}. Here, $Y_{23}$ and $Y_{34}$ denote the dimensionless jet-resolution scales associated with the transitions between the two- and three-jet configurations and between the three- and four-jet configurations, respectively, in the exclusive $e^+e^- k_t$ clustering algorithm~\cite{eekt}. This tail is typically associated with residual events containing hard gluon radiation. To preserve a clean two-jet topology, the jet-resolution parameters are further required to satisfy $Y_{23} \leq 0.1$ and $Y_{34} \leq 0.008$.

Finally, global kinematic and vertexing requirements are imposed to suppress residual backgrounds from ISR, multi-neutrino processes, and non-collision sources. Since ISR photons and escaping beam remnants are predominantly emitted along the beam direction, the missing-momentum direction is required to satisfy $|\cos\theta_{\text{miss}}| < 0.9$, which efficiently removes forward-peaked topologies. To further suppress multi-neutrino backgrounds, such as $W^+W^- \to \tau\nu \bar{q}q^\prime$, the missing-mass proxy $U_{\text{miss}} = E_{\text{miss}} - P_{\text{miss}}$ is used. Signal events with a single missing massless neutrino are expected to populate the region near $U_{\text{miss}} = 0$, whereas events with multiple missing particles extend to significantly larger values. The requirement $U_{\text{miss}} \leq 15$\,GeV is therefore applied. In addition, to reject cosmic-ray muons and decays in flight, the reconstructed muon track is required to be consistent with the primary interaction vertex by imposing $\chi^2_{\text{IP}} \equiv (d_0/\sigma_{d_0})^2 + (z_0/\sigma_{z_0})^2 < 10$. Here, $\sigma_{d_0}$ and $\sigma_{z_0}$ are the square roots of the corresponding diagonal elements of the track-parameter covariance matrix, evaluated for each track rather than treated as fixed detector-wide constants.

For illustration, for central tracks with $|\cos\theta|<0.5$, the median impact-parameter resolutions are approximately $\sigma_{d_0}=3.9\,\mu$m and $\sigma_{z_0}=4.1\,\mu$m for $5\leq p<10$\,GeV, improving to about $2.3\,\mu$m and $2.8\,\mu$m, respectively, for $20\leq p<50$\,GeV. These values illustrate the momentum dependence of the per-track impact-parameter uncertainties and are distinct from the intrinsic single-hit resolutions of the detector layers.

\begin{figure}[!htbp]
    \centering
    \includegraphics[width=0.8\textwidth]{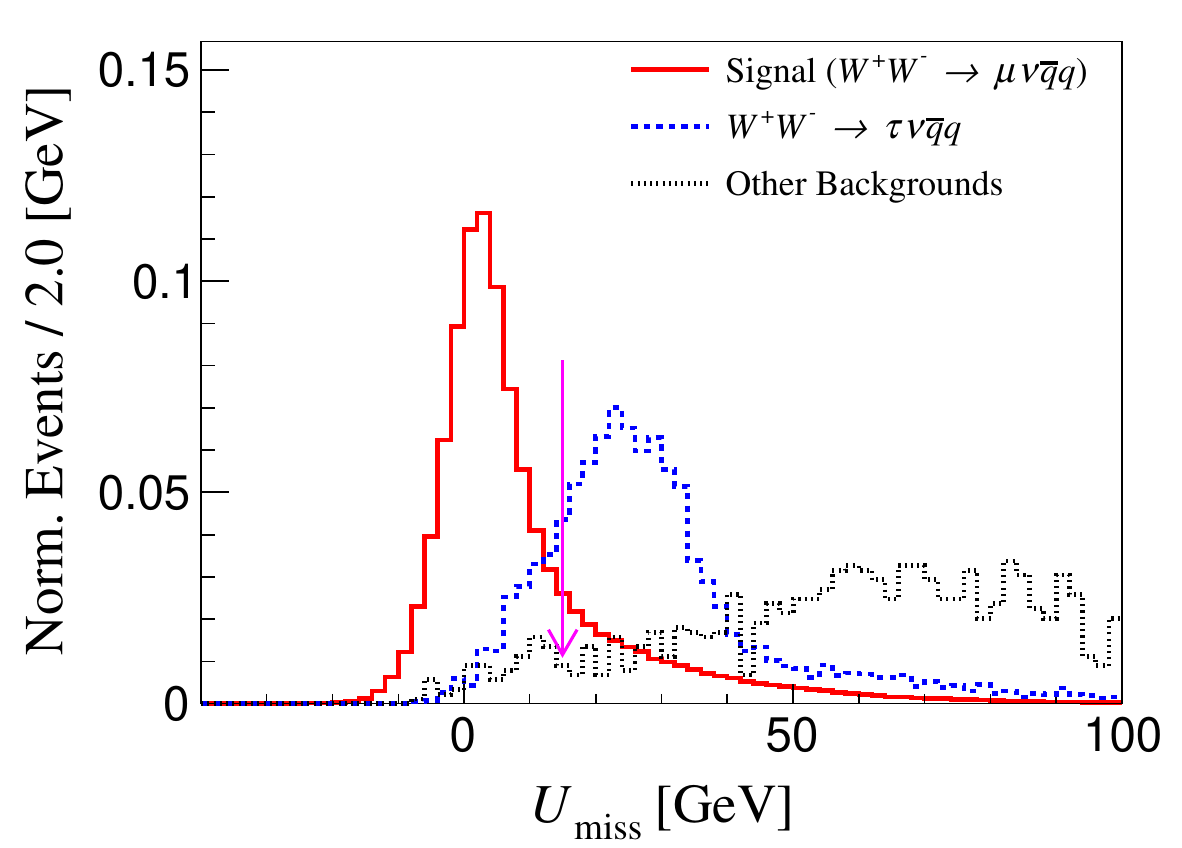}
    \caption{Normalized distributions of the missing-mass proxy variable, $U_{\text{miss}}$, after the $|\cos\theta_{\text{miss}}|$ requirement. The signal is shown by the red solid line, the $W^+W^- \to \tau\nu \bar{q}q^\prime$ background by the blue dashed line, and the remaining combined backgrounds by the black dotted line. The pink arrow indicates the requirement $U_{\text{miss}} \leq 15$\,GeV.}
    \label{fig:umiss_var}
\end{figure}

\subsection{Summary of the Selection}

The full selection is applied sequentially. Table~\ref{tab:cutflow} summarizes the cut flow for the representative signal channel $W^+W^-\to\mu\nu\bar{c}s$ and four background categories: $e^+e^-\to q\bar{q}$, $ZZ$, $W^+W^-\to\tau\nu\bar{q}q^\prime$, and $ZH$. The $e^+e^-\to q\bar{q}$ sample is retained to illustrate the suppression of the large two-fermion background, although its final yield is zero. The $W^+W^-\to e\nu\bar{q}q^\prime$ background was also simulated, but no events in the available MC sample survive the muon identification and isolation requirements. It is therefore omitted from Table~\ref{tab:cutflow}.

After the full selection, the final signal efficiencies and expected event yields for all six exclusive flavor channels are summarized in Table~\ref{tab:signal_yields}. The selection efficiency exceeds 53\,\% for the light- and charm-quark channels. The somewhat lower efficiencies for the $b$-quark channels, $W^+W^- \to \mu\nu \bar{u}b$ and $W^+W^- \to \mu\nu \bar{c}b$, are expected because semileptonic $b$-hadron decays can introduce additional neutrinos and soft muons inside the jets, which are occasionally rejected by the stringent $U_{\text{miss}}$ and isolation requirements.

Overall, the background suppression is highly effective. While the two-fermion and single-boson backgrounds are almost entirely removed, the residual background is dominated by the $W^+W^- \to \tau\nu \bar{q}q^\prime$ channel, with only small contributions from $ZZ$ and $ZH$. The final row of Table~\ref{tab:cutflow} gives 3,846 $ZZ$ events, 115,000 $W^+W^- \to \tau\nu \bar{q}q^\prime$ events, and 1,225 $ZH$ events, for a total residual background of 120,071 events. The $ZH$ entry combines 1,212 $ZH\to q\bar{q}H$ events and 13 $ZH\to\ell^+\ell^-H$ events. Relative to the inclusive selected signal yield, the background contamination is about 0.40\%, corresponding to an inclusive signal purity of about 99.6\%.

\begin{table}[htbp]
\centering
\caption{Summary of the sequential event selection for the representative signal channel $W^+W^- \to \mu\nu \bar{c}s$ and the dominant grouped background processes at $\sqrt{s}=240$\,GeV. The first and last rows list the expected yields normalized to 21.6\,ab$^{-1}$, while the intermediate rows list cumulative selection efficiencies.}
\label{tab:cutflow}
\scriptsize
\setlength{\tabcolsep}{3pt}
\renewcommand{\arraystretch}{1.1}
\begin{tabular}{>{\raggedright\arraybackslash}p{3.9cm}ccccc}
\toprule
\makecell[l]{Selection\\criterion} & \makecell[c]{$W^+W^- \to$\\$\mu\nu \bar{c}s$} & \makecell[c]{$e^+e^- \to$\\$q\bar{q}$}  & $ZZ$ & \makecell[c]{$W^+W^- \to$\\$\tau\nu \bar{q}q^\prime$} & $ZH$ \\
\midrule
Total events & 25,257,744 &1,168,708,160  & 12,020,184 & 57,499,848 & 4,399,256 \\
\midrule
$N_{\text{jet}} = 2$                                                    & 100.0\,\% & $99.9\,\%$  & 100.0\,\%  & 100.0\,\% & 93.1\,\%  \\
$5 \leq N_{\text{trk}} \leq 30$                                         & 99.0\,\%  & 94.9\,\%    & 97.6\,\%   & 99.6\,\%  & 47.8\,\%  \\
$N_{\ell} \geq 1$                                                       & 98.9\,\%  & 24.3\,\%    & 54.5\,\%   & 45.2\,\%  & 32.3\,\%  \\
$50 \leq \sum E_{\text{trk}} \leq 200$\,GeV                              & 97.8\,\%  & 19.2\,\%    & 44.7\,\%   & 42.0\,\%  & 31.0\,\%  \\
$P(\gamma_{\text{lead}}) \leq 60$\,GeV                                   & 97.4\,\%  & 15.5\,\%    & 43.2\,\%   & 41.8\,\%  & 30.9\,\%  \\ \midrule
$N_{\mu} \geq 1$                                                        & 97.4\,\%  & 6.6\,\%     & 36.7\,\%   & 19.4\,\%  & 16.8\,\%  \\
$P(\mu_{\text{lead}}) \geq 20$\,GeV                                      & 95.1\,\%  & 1.2\,\%     & 27.7\,\%   & 9.8\,\%   & 5.0\,\%   \\
$\sum P(\mu) \leq 100$\,GeV                                              & 90.5\,\%  & 1.2\,\%     & 10.0\,\%   & 9.8\,\%   & 5.0\,\%   \\
$80^\circ \leq \theta(\mu_{\text{lead}}, \text{di-jet}) \leq 170^\circ$ & 82.9\,\%  & 0.2\,\%     & 3.1\,\%    & 8.5\,\%   & 3.3\,\%   \\
$E_{\text{ratio}}(\mu_{\text{lead}}) \geq 0.9$                          & 79.7\,\%  & $<0.1\,\%$  & 1.8\,\%    & 8.2\,\%   & 2.3\,\%   \\ \midrule
$50 \leq M_{\text{di-jet}} \leq 120$\,GeV                                & 77.8\,\%  & $<0.1\,\%$  & 1.0\,\%    & 7.9\,\%   & 0.8\,\%   \\
$Y_{23} \leq 0.1$                                                       & 76.6\,\%  & $<0.1\,\%$  & 0.9\,\%    & 6.5\,\%   & 0.6\,\%   \\
$Y_{34} \leq 0.008$                                                     & 73.8\,\%  & $<0.1\,\%$  & 0.8\,\%    & 5.8\,\%   & 0.4\,\%   \\
$|\cos\theta_{\text{miss}}| \leq 0.9$                                   & 68.4\,\%  & 0.0\,\%     & 0.1\,\%    & 4.4\,\%   & 0.4\,\%   \\
$U_{\text{miss}} \leq 15$\,GeV                                           & 54.1\,\%  & 0.0\,\%     & $<0.1\,\%$ & 0.8\,\%   & $<0.1\,\%$ \\
$\chi^2_{\text{IP}} \leq 10$                                            & 53.8\,\%  & 0.0\,\%     & $<0.1\,\%$ & 0.2\,\%   & $<0.1\,\%$ \\
\midrule
Final expected yields                                                   & 13,588,666 &0   & 3,846    &    115,000   & 1,225     \\
\bottomrule
\end{tabular}
\renewcommand{\arraystretch}{1.0}
\setlength{\tabcolsep}{6pt}
\end{table}

The other simulated background processes have zero or negligible final yields at the precision shown. Since the selected events are not assigned to mutually exclusive reconstructed flavor categories, the residual background is included as a common component in the simultaneous template fit, rather than being assigned separately to the six signal channels.

\begin{table}[htbp]
\centering
    \caption{Final signal efficiencies and expected event yields for the six exclusive $W^+W^- \to \mu\nu \bar{q}_i q^\prime_j$ channels after the full selection at $\sqrt{s}=240$\,GeV with an integrated luminosity of 21.6\,ab$^{-1}$. The final background yields and their grouped composition are given in Table~\ref{tab:cutflow}; they are not assigned to individual signal channels because all selected events enter the simultaneous template fit.}
\label{tab:signal_yields}
\begin{tabular}{lcc}
\toprule
Signal Channel & Efficiency (\%) & Expected Yield \\
\midrule
$W^+W^- \to \mu\nu \bar{u}d$ & 58.2 & $1.475\times10^7$ \\
$W^+W^- \to \mu\nu \bar{u}s$ & 56.4 & $7.64\times10^5$    \\
$W^+W^- \to \mu\nu \bar{c}d$ & 55.2 & $7.48\times10^5$    \\
$W^+W^- \to \mu\nu \bar{c}s$ & 53.8 & $1.359\times10^7$ \\
$W^+W^- \to \mu\nu \bar{c}b$ & 44.7 & $2.10\times10^4$     \\
$W^+W^- \to \mu\nu \bar{u}b$ & 48.0 & $2.1\times10^2$        \\
\bottomrule
\end{tabular}
\end{table}
 
 
\section{Jet Flavor Tagging}

The precision of the CKM matrix element extraction relies critically on accurate identification of the quark flavors produced in hadronic $W$ decays. While heavy-flavor jets ($b$ and $c$) exhibit distinctive secondary vertices and mass signatures, discrimination among light-flavor jets ($u$, $d$, and $s$) remains experimentally challenging because of their similar fragmentation patterns. To address this problem, ParticleNet, a graph-neural-network-based jet flavor tagger, is employed for multi-class flavor classification.

The tagger is configured with four output categories: $b$, $c$, $s$, and $ud$.
The $u$ and $d$ jets are merged into a single light-flavor category, denoted as $ud$. For the targeted CKM measurement, an explicit separation between $u$ and $d$ jets is not required, since the relevant $W \to \bar{q}q^\prime$ channels can be distinguished mainly through the identification of $c$, $s$, and $b$ jets.  
ParticleNet is well suited to this task because it exploits constituent-level kinematic correlations by representing jets as particle clouds. Although more recent architectures such as ParT~\cite{qu2022particle} can yield modest performance improvements, ParticleNet is adopted here because it offers an attractive balance between tagging performance and computational cost for large-scale inference.

The implementation follows the general ParticleNet design but adopts a simplified architecture optimized for this analysis. Empirical tuning showed that the original, more complex configuration tended to overfit the training sample and did not provide a significant improvement in tagging performance. A lightweight model is therefore adopted, consisting of three EdgeConv blocks with 8 nearest neighbors and channel dimensions of (16, 16, 16), (32, 32, 32), and (64, 64, 64), respectively. The extracted features are aggregated through channel-wise global average pooling, followed by a fully connected layer with 64 units and ReLU activation. To further regularize the network and suppress overtraining, a dropout layer with a probability of 0.1 is applied before the final softmax output layer.

The model is trained with the Weaver framework~\cite{weaver_github}, which is based on PyTorch~\cite{pytorch}. The training sample contains one million simulated events for each of the four flavor categories, with an 80\,\%/20\,\% split between training and validation. The training is performed for 60 epochs with a batch size of 4,000 using the Adam optimizer~\cite{adam_optimizer} with an initial learning rate of 0.02.

To exploit the detector's precise tracking and fine-grained calorimetry, the input variables are chosen to provide a detailed description of each jet constituent. These features are grouped into three physically motivated categories:

\begin{itemize}
  \item \textbf{Particle kinematics}: These variables describe the spatial topology and energy flow of the jet. They include the Cartesian momentum components $(p_x, p_y, p_z)$, the logarithmic transverse momentum and energy ($\log p_T$ and $\log E$), and the jet-normalized variables $\log(p_T/p_T^{\text{jet}})$ and $\log(E/E^{\text{jet}})$, which characterize momentum sharing within the jet.

  \item \textbf{Displacement and vertexing}: These variables characterize the displacement of tracks from the primary interaction point and provide strong discrimination for $b$- and $c$-hadron decays because of their finite lifetimes. The inputs include the normalized transverse and longitudinal impact parameters, $d_0/\sigma_{d_0}$ and $z_0/\sigma_{z_0}$. The impact-parameter probability, $\exp(-\chi^2_{\text{IP}}/2)$, is also included, where $\chi^2_{\text{IP}} \equiv (d_0/\sigma_{d_0})^2 + (z_0/\sigma_{z_0})^2$ is assumed to follow a chi-squared distribution with two degrees of freedom.

  \item \textbf{Particle identification}: These variables exploit detector-response information to identify particle species and are particularly useful for tagging semileptonic heavy-flavor decays. The inputs include the reconstructed particle charge and mass. Two calorimeter-to-tracker observables, $E_{\text{em}}/|\mathbf{p}|$ and $(E_{\text{em}}+E_{\text{had}})/|\mathbf{p}|$, are also used, where $E_{\text{em}}$ and $E_{\text{had}}$ denote the associated electromagnetic and hadronic calorimeter energy deposits. These observables provide useful separation among electrons, muons, and hadrons.
\end{itemize}

\begin{figure}[!htb]
  \centering
  \includegraphics[width=0.8\textwidth]{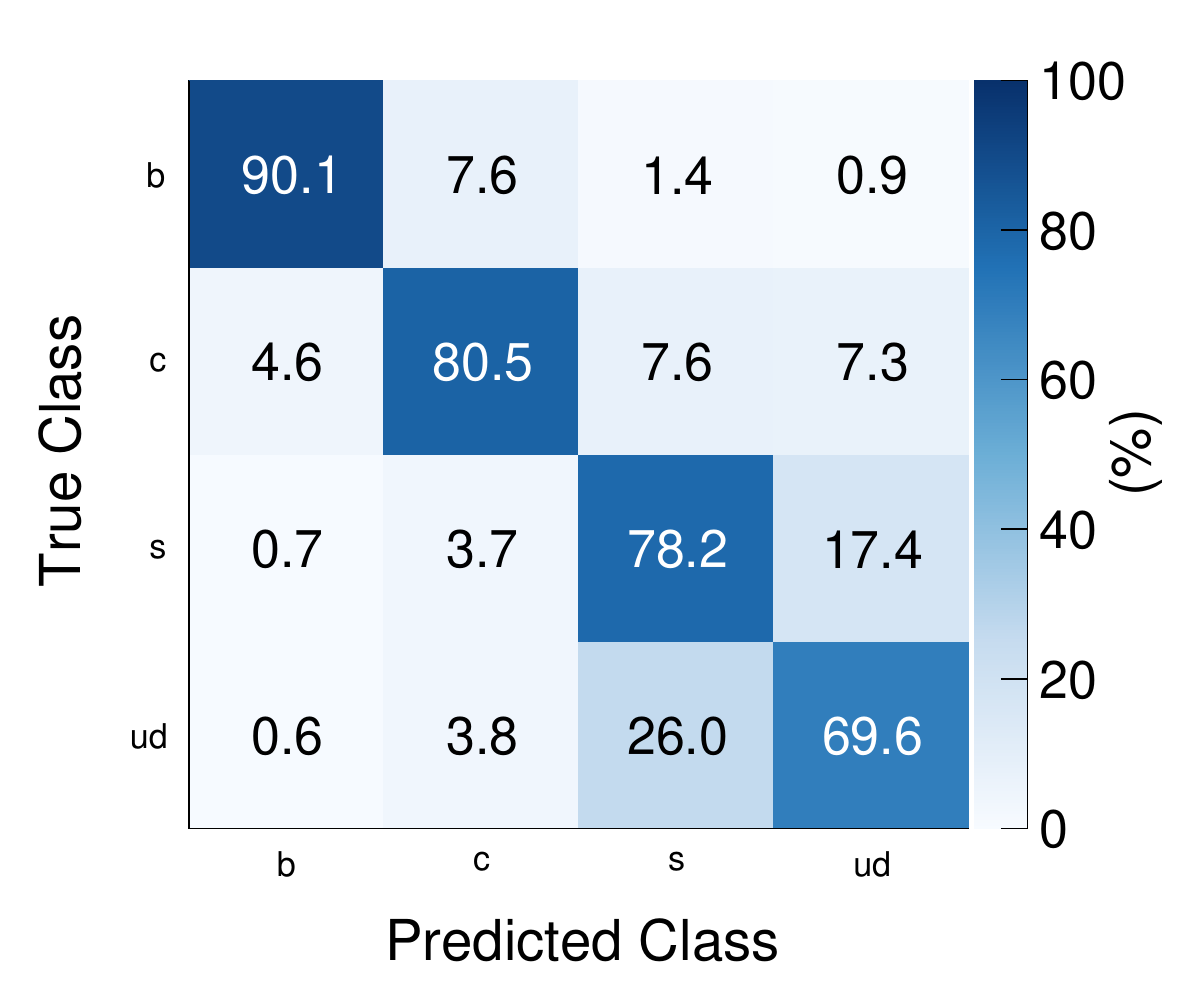}
  \caption{Confusion matrix of the ParticleNet model evaluated on the validation sample. Each matrix element gives the fraction of jets with a given true flavor (rows) assigned to a predicted flavor category (columns).}
  \label{fig:confusion_matrix}
\end{figure}

The performance of the lightweight ParticleNet model is evaluated on an independent validation sample. Figure~\ref{fig:confusion_matrix} shows the resulting confusion matrix, for which the overall multi-class classification accuracy reaches 79.6\,\%. As expected, the largest residual confusion occurs between the $s$ and $ud$ categories because of their similar detector signatures and the absence of distinctive secondary vertices. Nevertheless, the model provides strong discrimination between heavy-flavor ($b$, $c$) and light-flavor jets, together with useful $s$-tagging capability, which is sufficient for the downstream CKM analysis.

\begin{figure}[!htb]
  \centering
      \includegraphics[width=0.8\textwidth]{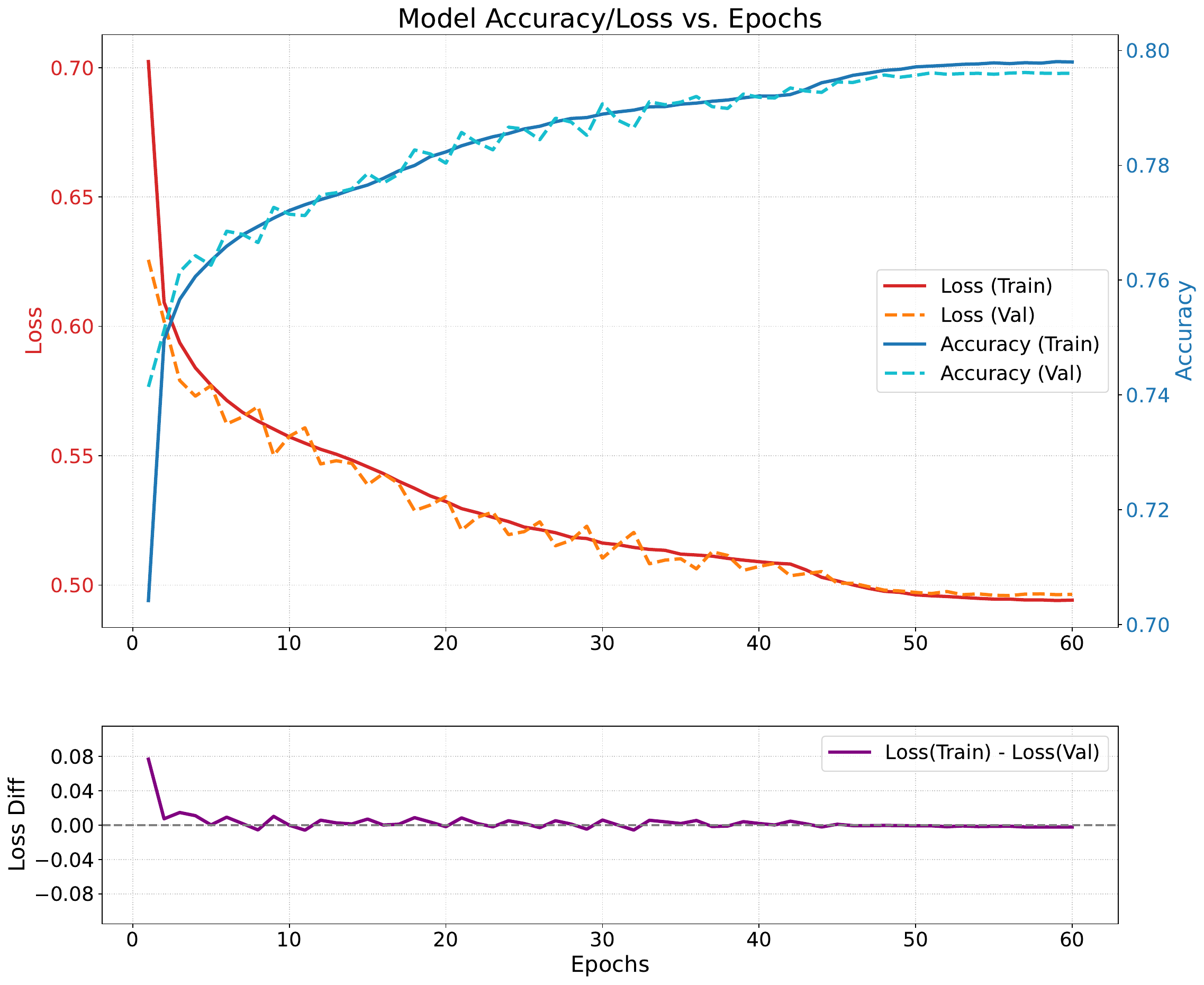}
  \caption{Training history of the ParticleNet model over 60~epochs. The upper panel shows the accuracy and loss for the training and validation samples. The lower panel shows the difference $\mathrm{Loss}(\mathrm{train})-\mathrm{Loss}(\mathrm{validation})$.}
  \label{fig:loss}
\end{figure}

The training history is shown in Fig.~\ref{fig:loss}. The training and validation curves exhibit stable convergence with no significant sign of overfitting. The best performance is reached after about 50 epochs, consistent with the chosen training configuration.

\section{CKM Matrix Element Extraction and Results}\label{sec:systematic}
 
The CKM matrix elements are extracted from the selected event sample using the jet flavor information provided by the jet flavor tagger. This section describes the construction of the flavor discriminants, the template fit used to determine the signal yields, and the propagation of statistical and systematic uncertainties to the extracted CKM matrix elements.

\subsection{$W$ branching fractions}

For each event with two reconstructed jets, the ParticleNet model provides a four-component probability vector for each jet,
\begin{equation}
\left(P_{i}(b), P_{i}(c), P_{i}(s), P_{i}(ud)\right), \quad \text{with}~\sum_{f}P_{i}(f)=1,
\end{equation}
where $i=1,2$ labels the jet and the probabilities are normalized to unity.

To construct an event-level discriminant for a flavor combination $h_1h_2$, the symmetrized raw score is defined as
\begin{equation}
S_{h_{1}h_{2}} \equiv P_{1}(h_{1}) P_{2}(h_{2}) + P_{1}(h_{2}) P_{2}(h_{1}),
\end{equation}
where $h_1$ and $h_2$ run over the four flavor categories $\{b,c,s,ud\}$. This definition accounts for the ambiguity in jet ordering, including the factor of two for the $ud$ combination. For the six $W$-decay flavor hypotheses considered in the analysis, $\mathcal{H}=\{cb,cd,cs,ud,ub,us\}$, the normalized discriminant for a specific signal channel is defined as
\begin{equation}
P_{ij} = \frac{S_{ij}}{\sum_{mn\in\mathcal{H}} S_{mn}}.
\end{equation}
The denominator therefore runs over the six $W$-decay flavor hypotheses considered in the analysis, and the six $P_{ij}$ values are normalized to unity. For a given signal channel, $P_{ij}$ tends to populate values close to unity, while background processes and mismatched flavor combinations are shifted toward smaller values.

\begin{figure}[!htb]
  \centering
  \includegraphics[width=0.48\textwidth]{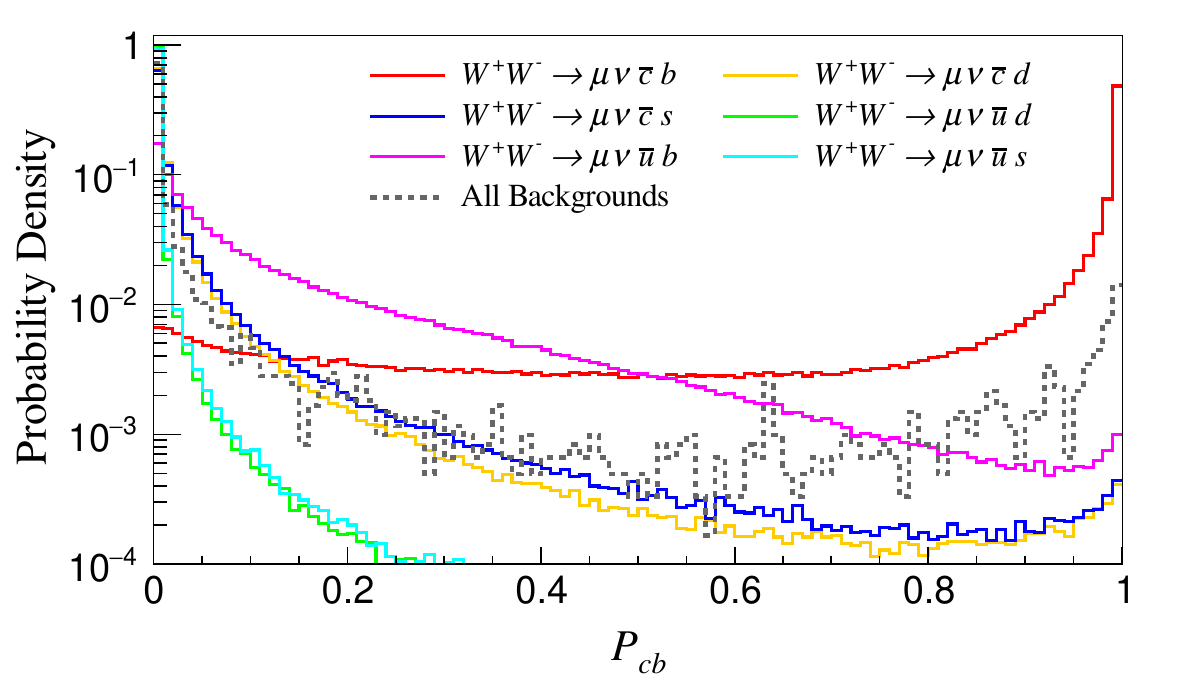}
  \includegraphics[width=0.48\textwidth]{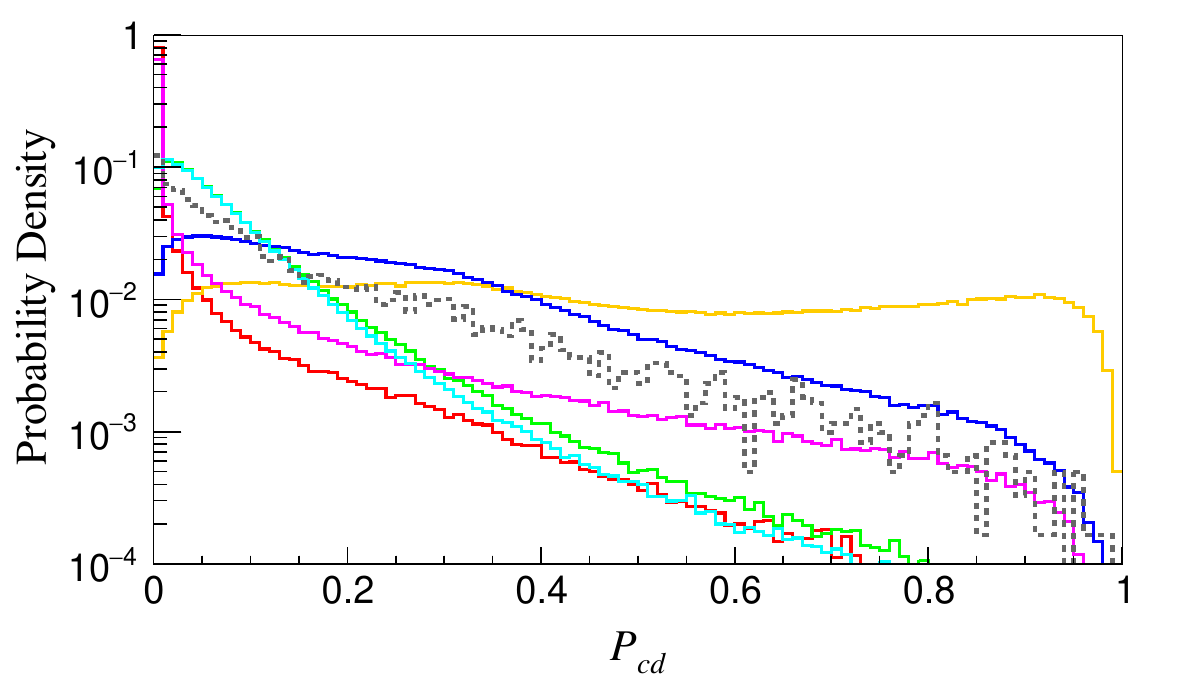}
  \includegraphics[width=0.48\textwidth]{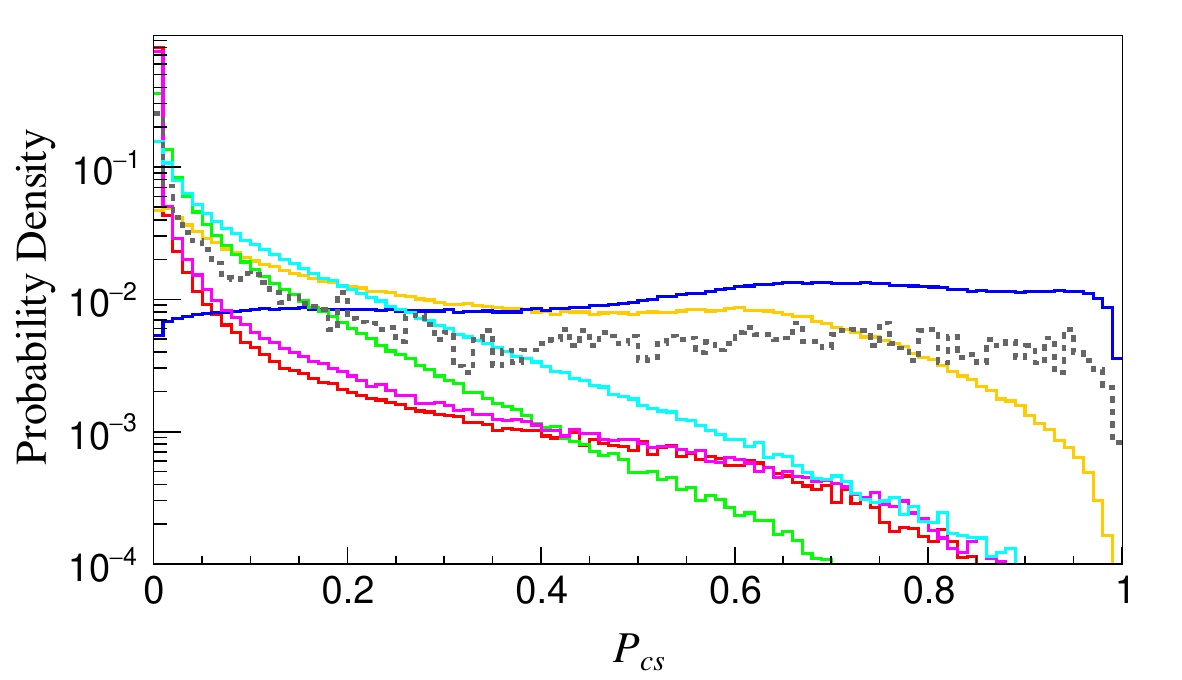}
  \includegraphics[width=0.48\textwidth]{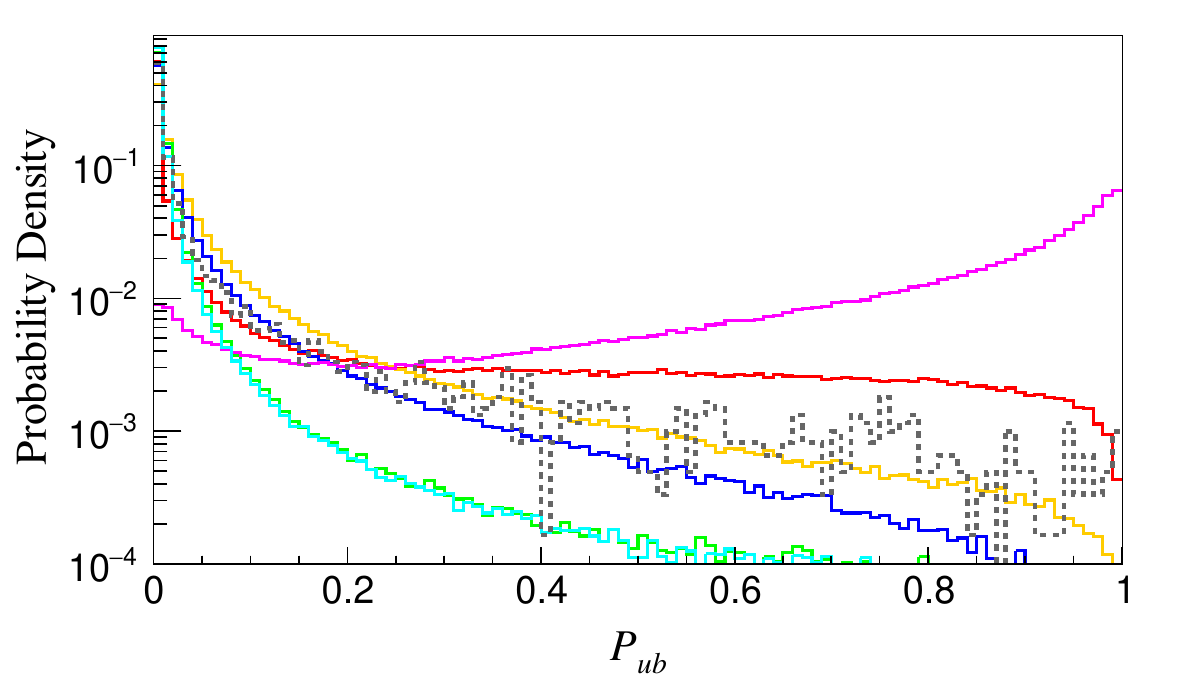}
  \includegraphics[width=0.48\textwidth]{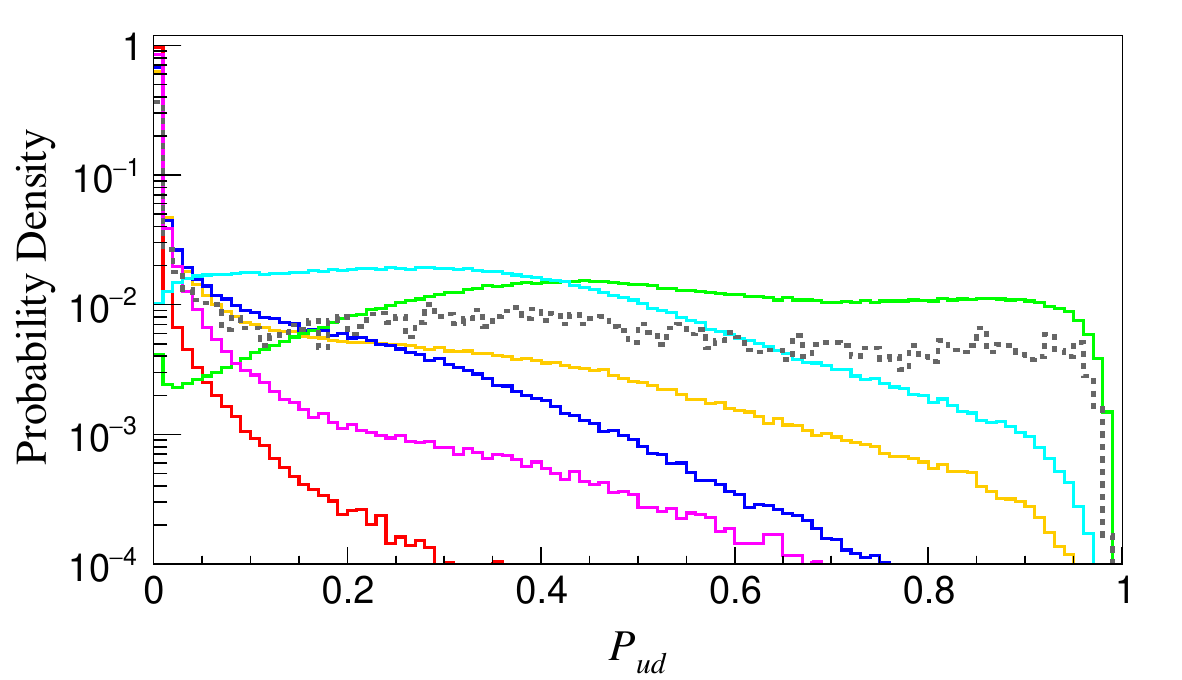}
  \includegraphics[width=0.48\textwidth]{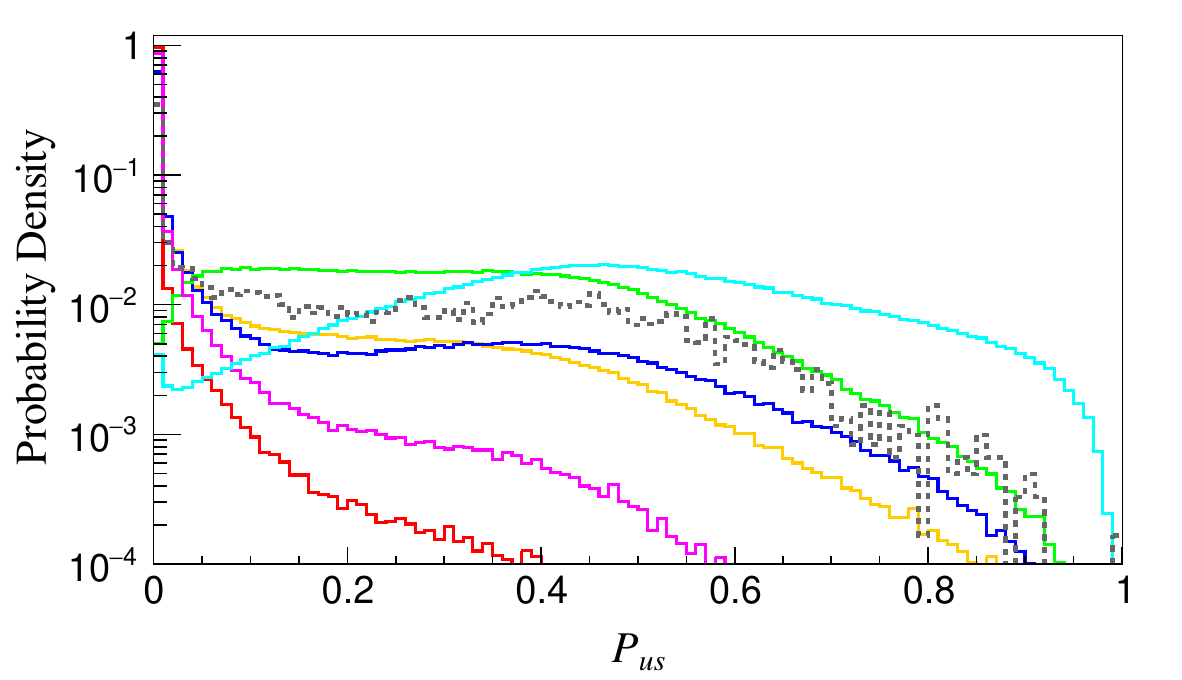}
    \caption{Event-level flavor-discriminant distributions for the six signal hypotheses, $P_{cb}$, $P_{cd}$, $P_{cs}$, $P_{ub}$, $P_{ud}$, and $P_{us}$. The solid curves show the signal distributions, while the dashed curves show the combined background contribution.}
    \label{fig:discriminants}
\end{figure}

The resulting event-level discriminant distributions are shown in Fig.~\ref{fig:discriminants}. The visible separation among the flavor hypotheses demonstrates that the ParticleNet output retains substantial discriminating power after being combined at the event level.

Given the large size of the selected sample, an unbinned fit would be computationally inefficient. The template fit is implemented as a binned extended maximum-likelihood fit to extract the yields of the different $W$ decays. In practice, the fit is performed by minimizing the negative log-likelihood (NLL) function,
\begin{equation}
-\ln \mathcal{L}(\mathbf{N}) = \sum_{k=1}^{N_{\text{bins}}} \left[ -n_k \ln\mu_k + \mu_k \right],
\label{eq:nll}
\end{equation}
where $n_k$ is the observed (or pseudo-data) event count in bin $k$, and $\mu_k$ is the corresponding expected number of events modeled as a linear superposition of normalized signal and background templates:
\begin{equation}
\mu_k = \sum_{ij} N_{ij} \mathcal{T}_{ij, k} + N_{\text{bkg}} \mathcal{T}_{\text{bkg}, k}.
\label{eq:mu_k}
\end{equation}
Here, $\mathbf{N} = \{N_{ij}\}$ denotes the set of free signal-yield parameters. The quantity $\mathcal{T}_{ij, k}$ represents the probability for an event from channel $ij$ to fall into bin $k$, as derived from the normalized Monte Carlo template for that channel. 
The $N_{\text{bkg}}$ and $\mathcal{T}_{\text{bkg}, k}$ denote the expected yield and the corresponding normalized template shape of the total residual background, respectively. The background yield is fixed to its nominal SM prediction, while the background shape is fixed to the simulated template. Since the likelihood is Poisson in the observed bin counts, the statistical fluctuations of the selected sample, including its background component, are propagated to the fitted signal-yield uncertainties. The uncertainty in the fixed background normalization is not included in the statistical uncertainty.
Because of its extremely small expected yield, the $\bar{u}b$ component ($N_{ub}$) is fixed. The free parameters of the fit are therefore the five remaining signal yields: $N_{cb}$, $N_{cd}$, $N_{cs}$, $N_{ud}$, and $N_{us}$.

The branching fractions are obtained from the fitted signal yields through
\begin{equation}
{\rm BR}(W^- \to \bar{q}_i q^\prime_j) = \frac{N_{ij}}{\varepsilon_{ij} \cdot N_{\mu\nu \bar{q}q^\prime}},
\end{equation}
where $\varepsilon_{ij}$ denotes the total selection efficiency for channel $ij$ and $N_{\mu\nu \bar{q}q^\prime}$ is the total number of inclusive $W^+W^- \to \mu\nu \bar{q}q^\prime$ events. Using the relation ${\rm BR}(W^- \to \bar{q}_i q^\prime_j) = \frac{1}{2}|V_{ij}|^2$, the fitted branching fractions can be translated directly into the magnitudes of the CKM matrix elements.

\begin{figure}[!htb]
  \centering
  \includegraphics[width=0.48\textwidth]{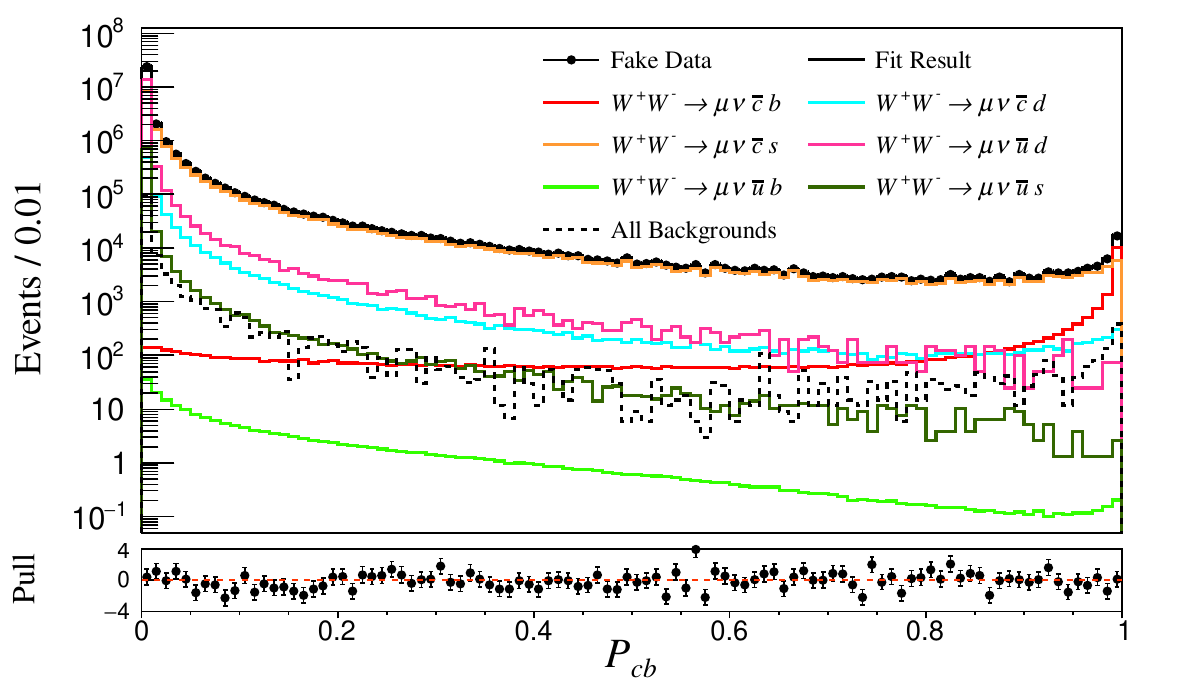}
  \includegraphics[width=0.48\textwidth]{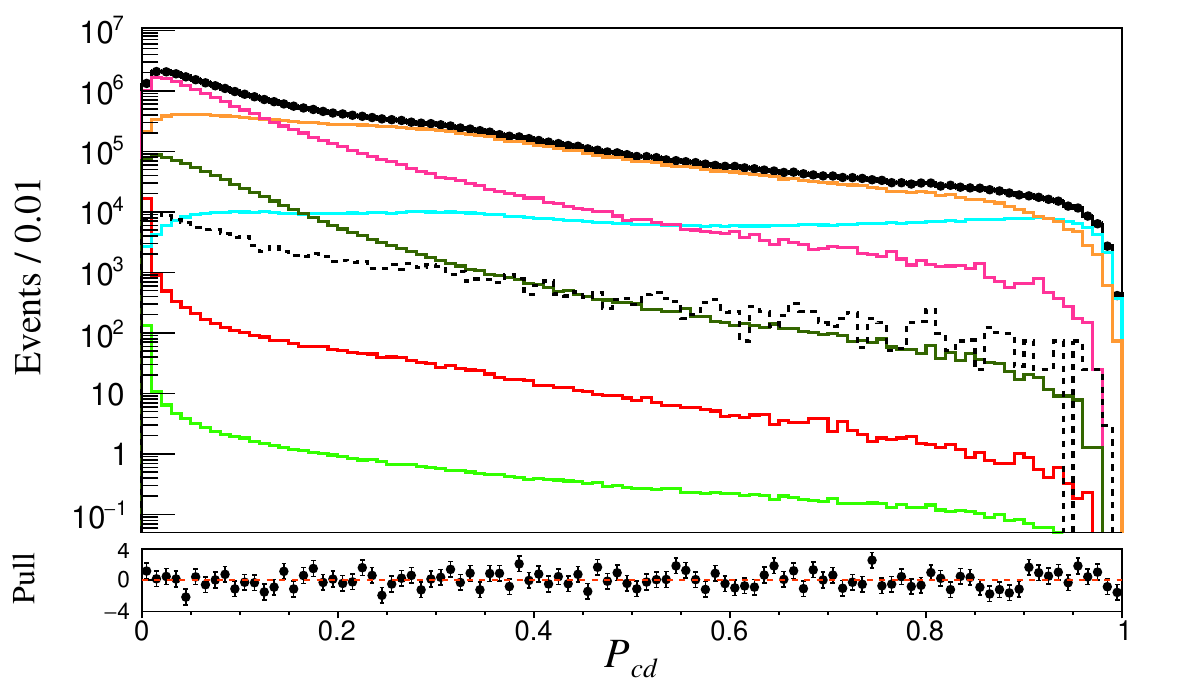}
  \includegraphics[width=0.48\textwidth]{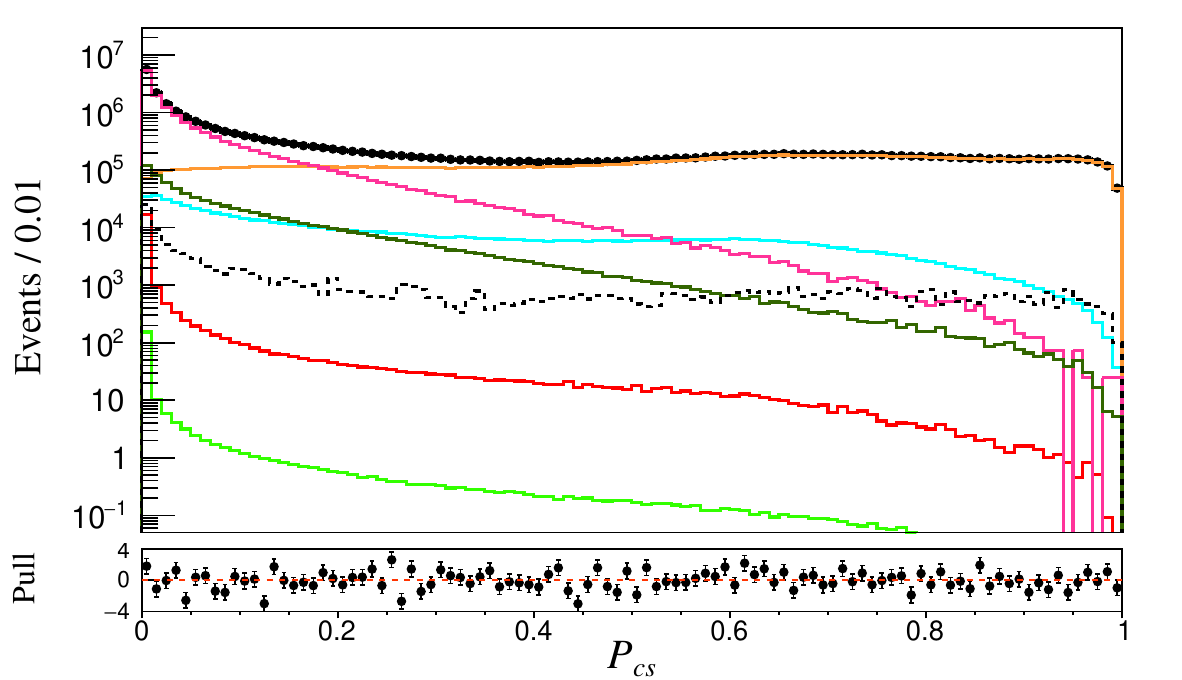}
  \includegraphics[width=0.48\textwidth]{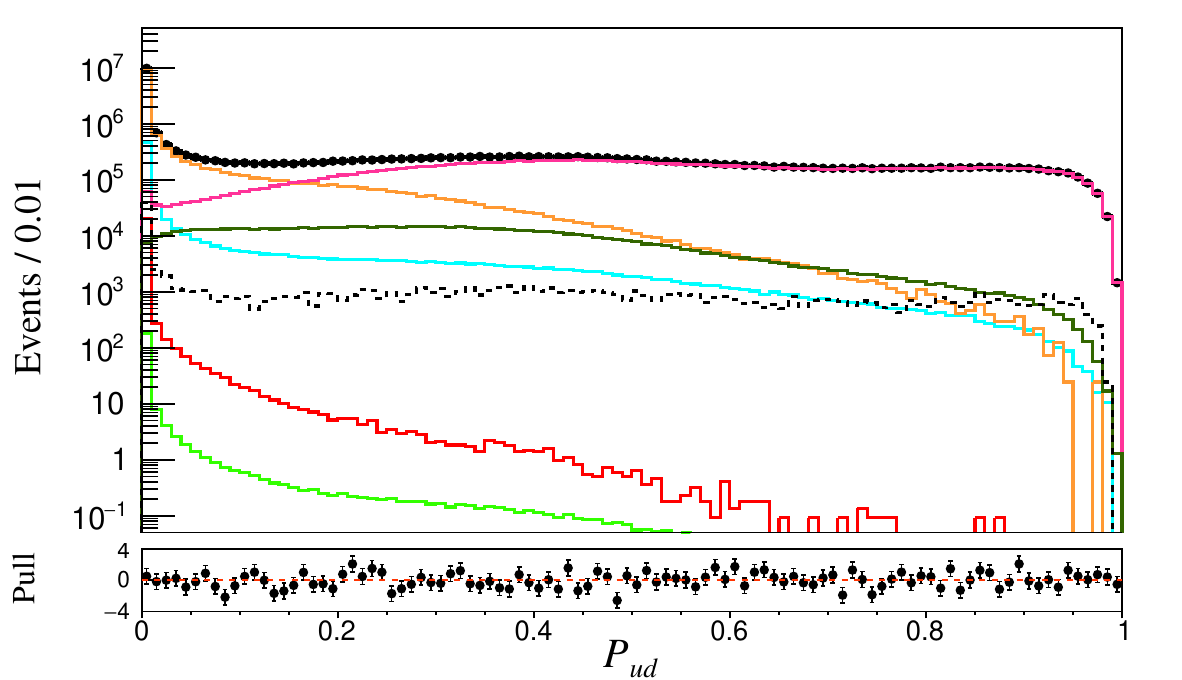}
  \includegraphics[width=0.48\textwidth]{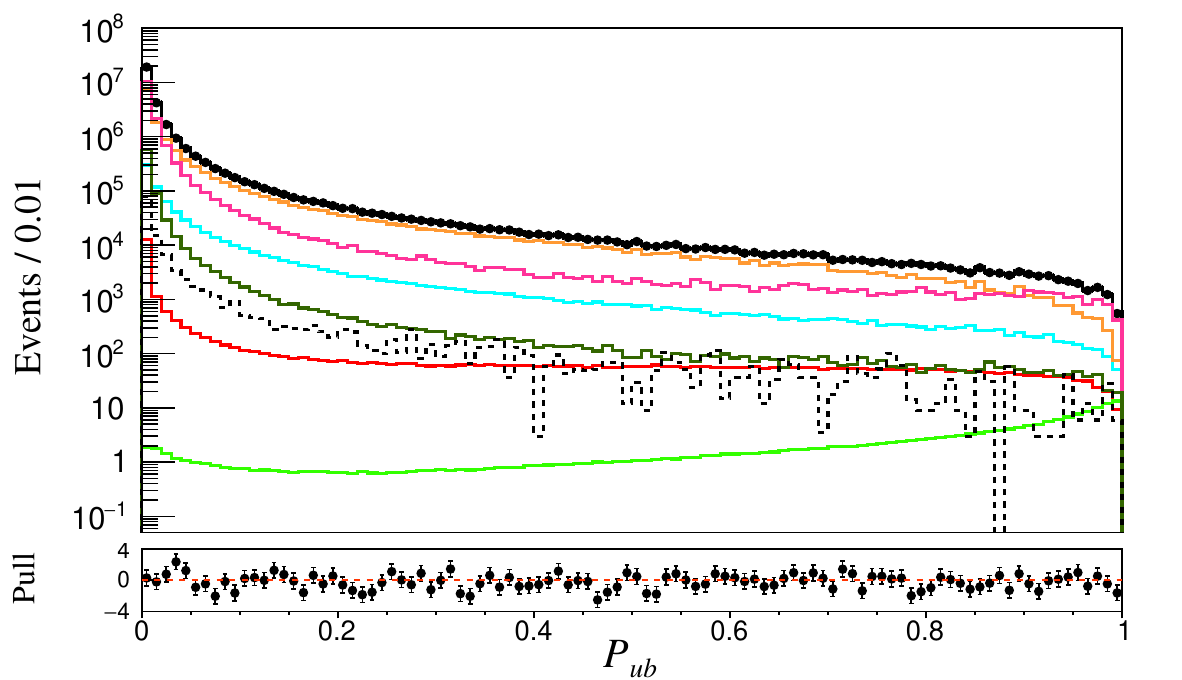}
  \includegraphics[width=0.48\textwidth]{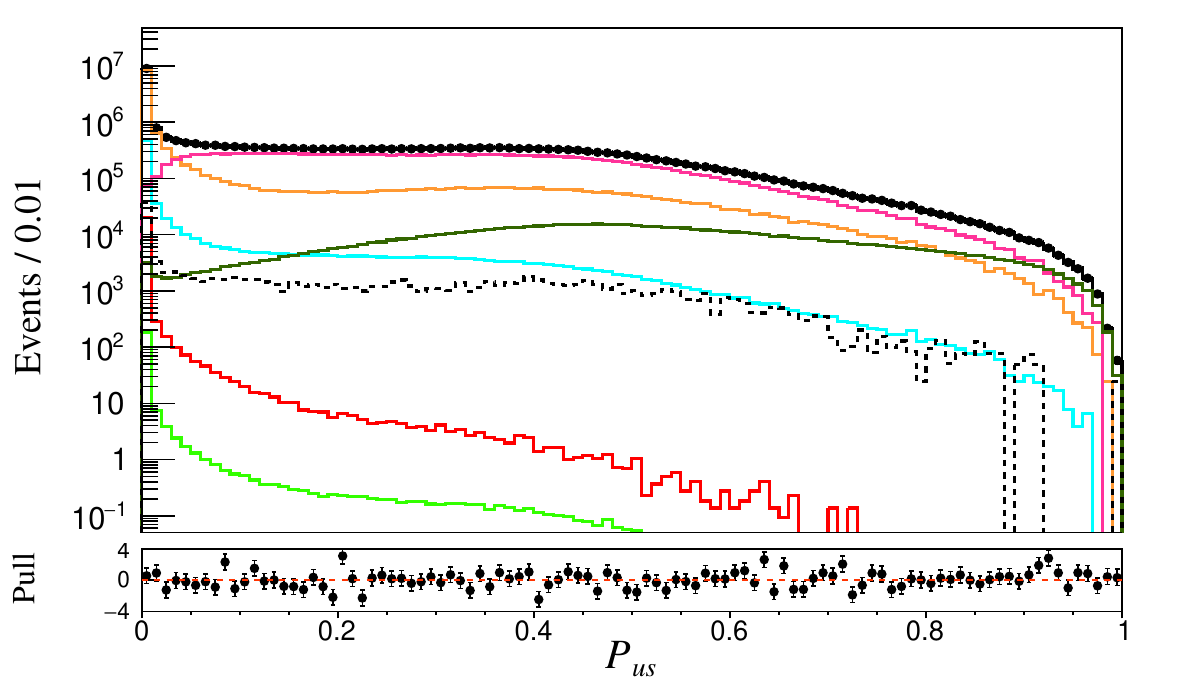}
  \caption{Projections of the template fit onto the event-level flavor discriminants for the six channels $W^- \to \bar{c}b$, $\bar{c}d$, $\bar{c}s$, $\bar{u}d$, $\bar{u}b$, and $\bar{u}s$, shown from left to right and from top to bottom. For each channel, the upper subpanel shows the pseudo-data as points with error bars, the total fit model as a solid black line, the individual signal components as colored solid lines, and the combined background contribution as a dashed line. The corresponding lower subpanel shows the pull distribution.}
  \label{fig:fit_results}
\end{figure}

The fit results are summarized in Table~\ref{tab:fit_results}, and the corresponding projections are shown in Fig.~\ref{fig:fit_results}. 
The fitted yields are in good agreement with the expected selected yields.

\begin{table}[htbp]
\centering
\caption{Expected selected yields, fitted signal yields $N_{ij}$, and the corresponding relative uncertainties obtained from the template fit.}
\label{tab:fit_results}
\begin{tabular}{lccc}
\toprule
\textbf{$N_{ij}$} & \textbf{Expected selected yield} & \textbf{Fitted yield} & \textbf{Relative uncertainty} \\
\midrule
$N_{cb}$ & $2.105 \times 10^4$ & $(2.123 \pm 0.025) \times 10^4$ & 1.18\,\% \\
$N_{cd}$ & $7.477 \times 10^5$ & $(7.454 \pm 0.027) \times 10^5$ & 0.36\,\% \\
$N_{cs}$ & $1.3589 \times 10^7$ & $(1.3588 \pm 0.0004) \times 10^7$ & 0.02\,\% \\
$N_{ud}$ & $1.4746 \times 10^7$ & $(1.4751 \pm 0.0004) \times 10^7$ & 0.02\,\% \\
$N_{us}$ & $7.640 \times 10^5$ & $(7.614 \pm 0.031) \times 10^5$ & 0.41\,\% \\
\bottomrule
\end{tabular}
\end{table}

\subsection{Uncertainties for CKM Matrix Elements}

According to Eq.~\eqref{eq:br}, the CKM matrix elements are extracted from the fitted yields using 
\begin{equation}
|V_{ij}| = \sqrt{\frac{2 N_{ij}}{\varepsilon_{ij} \cdot N_{\mu\nu \bar{q}q^\prime}}}.
\end{equation}

\textbf{Statistical uncertainties:}

For each channel, the relative statistical uncertainty on $|V_{ij}|$ follows directly from the fitted yield uncertainty,
\begin{equation}
\frac{\delta|V_{ij}|}{|V_{ij}|} = \frac{1}{2} \frac{\delta N_{ij}}{N_{ij}},
\end{equation}
where $\delta N_{ij}$ is the statistical uncertainty returned by the template fit as listed in Table~\ref{tab:fit_results}.

\textbf{Systematic uncertainties:}

\begin{itemize}

    \item \textbf{Detector resolution effects:}
    Detector-related uncertainties are evaluated by varying the tracking spatial resolutions and calorimeter energy resolutions in the Delphes simulation by $\pm 10\,\%$. For each of the six signal channels, the relative change in selection efficiency is calculated. For each detector source, the absolute relative efficiency changes are weighted by the expected selected signal yields, and the larger of the $+10\,\%$ and $-10\,\%$ variations is taken as the envelope. The resulting tracking and calorimeter contributions are approximately $0.14\,\%$ and $0.11\,\%$, respectively, at the selection-efficiency level. Treating them as independent gives a common, yield-weighted efficiency-level detector uncertainty of approximately $0.18\,\%$ in quadrature. Since $|V_{ij}|\propto\epsilon_{ij}^{-1/2}$, this corresponds to a relative detector-related uncertainty of approximately $0.09\,\%$ on the extracted CKM matrix elements.

    \item \textbf{Parton-shower and hadronization modeling:}
    The event-level flavor discriminants depend on the modeling of hadronic $W$ decays. Final-state QCD radiation, ISR, and hadronization can modify the reconstructed jet kinematics and therefore the jet flavor tagging response. 
    A previous study estimated a related jet flavor tagging and modeling uncertainty of about 0.8\,\% by comparing different Monte Carlo generators~\cite{Liang_Vcb_HiggsFactory}.
    In a full analysis, this uncertainty could be further reduced, potentially toward the 0.1\,\% level, by validating the jet fragmentation and the jet flavor tagging response with high-statistics control samples, in particular from the $Z$ pole, and by incorporating the resulting constraints into the signal modeling.

    \item \textbf{Perturbative corrections to the $W$ partial width:}
    The extraction of CKM matrix elements from hadronic $W$ branching fractions uses the perturbative expression for the partial width, including QCD, electroweak, and mixed electroweak--QCD corrections, as shown in Eq.~\eqref{eq:partial_width}. These correction factors are flavor independent to a very good approximation and therefore largely cancel in the normalized branching-fraction relation used in Eq.~\eqref{eq:br}. The residual uncertainty from missing higher-order terms and possible small flavor-dependent effects is expected to be subdominant for the present projection. It is therefore not included as a separate numerical uncertainty in this study, but should be revisited in a full precision analysis with updated perturbative calculations and input parameters.

\end{itemize}

The projected precision is summarized in Table~\ref{tab:ckm_results}. For completeness, the table also includes the current PDG direct-average precisions and representative single-experiment benchmarks for the five CKM elements considered here.
The most prominent improvement is obtained for $|V_{cs}|$, which benefits from both the large $W\to \bar{c}s$ branching fraction and the improved charm--strange flavor separation, reaching a projected statistical precision of $0.01\,\%$. Even with the common, yield-weighted detector-related relative uncertainty on $|V_{ij}|$ of approximately $0.09\,\%$, this precision would remain substantially better than the current PDG direct-average precision. The projected $0.59\,\%$ statistical precision for $|V_{cb}|$ would provide a direct electroweak determination from hadronic $W$ decays, independent of semileptonic $B$-decay form-factor inputs. A sizable improvement is also expected for $|V_{cd}|$, for which the projected statistical precision reaches $0.18\,\%$, about an order of magnitude better than the current PDG direct-average precision and substantially better than existing single-experiment determinations.

\begin{table}[htbp]
    \centering
    \caption{Projected precision for CKM matrix element measurements at the CEPC with an integrated luminosity of 21.6\,ab$^{-1}$. The statistical precision from the template fit and the common, yield-weighted CKM-element-level detector-related systematic uncertainty are listed separately, together with the corresponding PDG direct-average precision and representative single-experiment benchmarks.}
    \label{tab:ckm_results}
    \small
    \begin{tabular}{lcccp{4.2cm}}
    \toprule
    \multirow{2}{*}{$|V_{ij}|$} & \multicolumn{2}{c}{\textbf{CEPC}} & \multirow{2}{*}{\textbf{PDG direct avg.}} & \multirow{2}{*}{\textbf{Single-exp.}} \\
    \cmidrule(lr){2-3}
     & \textbf{Stat.} & \textbf{Det. syst.} &  &  \\
    \midrule
    $|V_{ud}|$ & 0.01\,\% & \multirow{5}{*}{$\sim 0.09\,\%$} & 0.03\,\% & PIBETA: 0.28\,\%~\cite{PIBETA_Vud,pdg2024} \\
    $|V_{us}|$ & 0.20\,\% &   & 0.38\,\% & KLOE+lattice: 0.18\,\%~\cite{KLOE_Vus,pdg2024} \\
    $|V_{cd}|$ & 0.18\,\% &   & 1.81\,\% & BESIII: 3.4\,\%~\cite{BESIII_Vcd} \\
    $|V_{cs}|$ & 0.01\,\% &   & 0.62\,\% & BESIII: 1.7\,\%~\cite{BESIII_Vcs} \\
    $|V_{cb}|$ & 0.59\,\% &   & 2.92\,\% & Belle-based: 1.5\,\%~\cite{Belle_Vcb,pdg2024} \\
    \bottomrule
    \end{tabular}
    \end{table}
\section{Summary and Conclusions}\label{sec:summary}

This study evaluates the CEPC potential for precision measurements of CKM matrix elements through semileptonic $W$-boson decays at $\sqrt{s}=240$\,GeV. Assuming an integrated luminosity of 21.6\,ab$^{-1}$, a simultaneous extraction of $|V_{ud}|$, $|V_{us}|$, $|V_{cd}|$, $|V_{cs}|$, and $|V_{cb}|$ is performed using the template fit. The analysis combines a dedicated event selection with ParticleNet-based jet flavor tagging to separate the relevant hadronic $W$ decay modes. This approach achieves a high signal purity after event selection and provides effective discrimination among the exclusive hadronic $W$ decay channels.

As summarized in Table~\ref{tab:ckm_results}, the projected sensitivities show that the large CEPC $W^+W^-$ dataset can turn hadronic $W$ decays into a statistically powerful probe of CKM structure. The most significant improvement is expected for $|V_{cs}|$, where the large $W\to \bar{c}s$ branching fraction and effective jet flavor tagging for charm and strange jets lead to a statistical precision substantially better than current direct measurements. The projected sensitivities to $|V_{cb}|$ and $|V_{cd}|$ are also particularly relevant: the former would provide a direct electroweak-boson determination independent of semileptonic $B$-decay form-factor inputs, while the latter would substantially improve upon existing direct measurements. For $|V_{ud}|$ and $|V_{us}|$, the measurement would not replace the most precise low-energy determinations, but it would offer an independent cross-check based on different experimental and theoretical systematics.

More broadly, this study indicates that the CEPC could establish a dedicated future program of direct CKM measurements using on-shell $W$ decays. Its large $W^+W^-$ sample, clean collision environment, and strong jet flavor tagging capabilities for both heavy- and light-flavor jets make it particularly well suited to this task. The use of machine-learning-based jet flavor tagging, exemplified here by ParticleNet, is a central ingredient of this program because it enables multi-class flavor separation beyond traditional heavy-flavor jet tagging. The systematic uncertainties considered here are preliminary and conservative, especially those associated with detector-performance variations and jet flavor tagging calibration. Further progress will rely on improved control of jet flavor tagging calibration and mistag rates, QCD radiation and hadronization modeling, and background normalization. Overall, the CEPC offers a realistic opportunity to turn hadronic $W$ branching fractions into precision probes of CKM structure in a way that is complementary to both low-energy flavor measurements and existing hadron-collider results.

\bmhead{Acknowledgements}
This work is supported in part by the Innovation Project of the Institute of High Energy Physics, Chinese Academy of Sciences and the Program of Science and Technology Development Plan of Jilin Province of China under contract No.~20230101021JC.

\section*{Statements and Declarations}

\noindent\textbf{Data availability}

The datasets generated and analyzed during this study
are available from the corresponding author upon
reasonable request.

\medskip
\noindent\textbf{Competing interests}

The authors declare that they have no competing interests.


\end{document}